\documentclass[11pt,titlepage,a4paper]{report}

\usepackage{amsfonts}
\usepackage{amssymb}
\usepackage{amsthm}
\usepackage{latexsym,amsmath}
\usepackage{graphicx}
\usepackage{fancyvrb}
\usepackage{float}
\usepackage{appendix}
\usepackage{titlepic}

\usepackage{tikz}
\usetikzlibrary{trees}
\tikzstyle{every node}=[draw=black,thick,anchor=west]
\tikzstyle{selected}=[draw=red,fill=red!30]
\tikzstyle{optional}=[dashed,fill=gray!50]

\newcommand{\PP}{\mathbb{P}}

\DeclareMathOperator*{\argmin}{\arg\!\min}
\DeclareMathOperator*{\argmax}{\arg\!\max}

\DeclareMathOperator{\di}{d\!}
\newcommand*\Bell{\ensuremath{\boldsymbol\ell}}

\newcommand{\Pm}{\mathbb{P}}
\newcommand{\simiid}{\stackrel{iid}{\sim}}

\newcommand{\Var}{\mathrm{Var}}
\newcommand{\RE}{\mathrm{RE}}

\newcommand{\e}{\text{e}}

\newcommand{\bX}{\mathbf{X}}
\newcommand{\bx}{\mathbf{x}}
\newcommand{\by}{\mathbf{y}}
\newcommand{\bY}{\mathbf{Y}}
\newcommand{\bz}{\mathbf{z}}
\newcommand{\bZ}{\mathbf{Z}}
\newcommand{\I}{\mathbb{I}}

\newcommand{\Em}{\mathbb{E}}

\newcommand{\reff}[2]{\ref{#1}.\ref{#2}}

\newcounter{Stuff}[section]
\newenvironment{Algorithm}{\medskip
                            \refstepcounter{Stuff}
                            \noindent
                           {\bf \small Algorithm \thesection.\theStuff.}\ }
                           {\medskip}
\newenvironment{Definition}{\medskip
                            \refstepcounter{Stuff}
                            \noindent
                           {\bf \small Definition \thesection.\theStuff.}\ }
                           {\medskip}

\newenvironment{Theorem}{\medskip
                            \refstepcounter{Stuff}
                            \noindent
                           {\bf \small Theorem \thesection.\theStuff.}\
													 \begingroup \sl} 
                           {\endgroup\medskip}

\newenvironment{Lemma}{\medskip
                            \refstepcounter{Stuff}
                            \noindent
                           {\bf Lemma \thesection.\theStuff.}\ %
                            \begingroup \sl}
                           {\endgroup\medskip}

\newenvironment{Example}{\medskip
                            \refstepcounter{Stuff}
                            \noindent
                           {\bf \small Example \thesection.\theStuff.}\ }
                           {\qed}
													{}

\titlepic{\includegraphics[scale=.8]{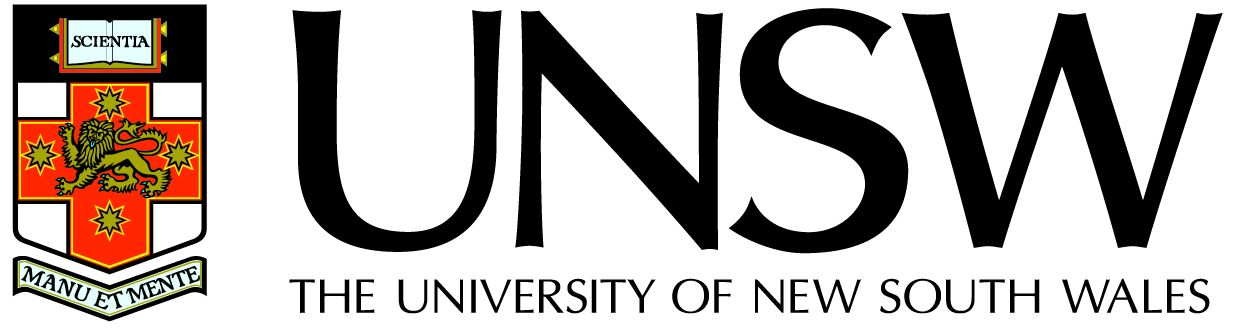}}

\title{Rare-event Probability Estimation\\ via Empirical Likelihood Maximization}

\author{Alvin Huang\\
Supervisor: Dr. Zdravko Botev\\
\\\\
Submitted in partial fulfillment of the\\
                requirements of the degree of \\
                {\sc Bachelor of Science  Honours},\\  {\sc School of Mathematics and Statistics},\\
								{\sc University of New South Wales}}

\begin{document}

\maketitle

\section*{Acknowledgements}

I would like to sincerely thank my supervisor Dr.~Zdravko~Botev for his patient guidance, advice and tutelage he has provided throughout my Honours year.
I am grateful for him taking time to share his interesting research ideas that has perked my interests.  
Also, I must express my gratitude to my colleagues Aaron and Victor for keeping me company and taking the same journey in 2013.
Discussions with them are insightful and helped me keep on track.
I would also like to thank the School of Mathematics and Statistics at UNSW for the fruitful experience over the last 5 years.   
Finally, I would like to thank my family for their care and support.  

\begin{abstract}

We explore past and recent developments in rare-event probability estimation with a particular focus on a novel Monte Carlo technique -- Empirical Likelihood Maximization (ELM). This is a versatile method that involves sampling from a sequence of densities using MCMC and maximizing an empirical likelihood. The quantity of interest, the probability of a given rare-event, is estimated by solving a convex optimization program related to likelihood maximization. Numerical experiments are performed using this new technique and benchmarks are given against existing robust algorithms and estimators. 
\paragraph{Keywords:} Rare-event probability estimation; Monte Carlo methods; Importance sampling; Markov chain Monte Carlo; Empirical Likelihood; Convex Optimization

 \end{abstract}




%
%



\chapter{Introduction}\label{c-introduction}

One of the hallmark problems in Monte Carlo simulation is the efficient estimation of a high-dimensional integral with the form
\begin{align}
\label{eq:Z_integral}
\mathcal{Z} = \int f(\bx) H(\bx)  \di\bx=  \Em_f H(\bX),
\end{align}
where $H:\mathbb{R}^d \rightarrow \mathbb{R}$ is an arbitrary real-valued function and $\bX=(X_1,\dots,X_d)$ is a $d$--dimensional random vector with joint probability density function $f$. 

These integrals arise in various contexts such as in Bayesian statistics,  insurance risk, financial mathematics, statistical mechanics, queuing analysis, reliability theory, and combinatorial problems in computer science (see~\cite{asmussen2007stochastic,kroese2011handbook}). An important class of estimation problems with the form \eqref{eq:Z_integral} is rare-event probability estimation. For this particular class of problems, $H$ is the indicator function
\begin{align*}
H(\bx) =\I\{S(\bx)\geq \gamma \}=
\begin{cases}
 \ 1 & \text{if} \ S(\bx) \geq \gamma \\
 \ 0 & \text{if} \ S(\bx) < \gamma,
\end{cases}
\end{align*}
where $S$ is a real-valued function and  $\gamma$ is a level or threshold parameter. That is, $H$ represents the occurrence of some given event. Define $\ell$ to be the probability of this event, then 
\begin{equation}\label{eq:rare_probability_integral}
\ell= \int f(\bx) \I\{S(\bx)\geq \gamma \}  \di\bx= \Em_f \I\{S(\bx)\geq \gamma \} = \mathbb{P}_f(S(\bX)\geq\gamma). 
\end{equation}

Of interest is when the probability $\ell$ happens to be very small,  say, with order less than $10^{-4}$. In such cases, we define the event $\{S(\bX) \geq \gamma  \}$ to be a \emph{rare-event} and $\ell$ to be a \emph{rare-event probability}. For example, $\ell$ may represent the probability of ruin for an insurance company such that $S(\bX)$ is the aggregate claims received and $\gamma$ is their current capital level. Other examples include telecommunications where $\ell$ the probability of buffer overflow, or in reliability theory where $\ell$ is the probability of failure before some time $\gamma$.

These probabilities are represented by high-dimensional integrals whose integrands are non-smooth. In effect, numerical integration techniques such as traditional quadrature rules and quasi-Monte Carlo methods are typically inapplicable. Thus, we resort to Monte Carlo simulation, whose convergence is usually unaffected by the dimensions of the integration and the smoothness of the integrand.

In this thesis, we utilize Monte Carlo methods to estimate these high-dimensional integrals. 
Particular focus is placed on a novel Monte Carlo technique, Empirical Likelihood Maximization (ELM). This is a versatile method that involves sampling from a sequence of densities using Markov chain Monte Carlo (MCMC) and maximizing an empirical likelihood.
The (rare-event) probability of interest is estimated by solving a convex optimization program similar to maximizing a likelihood function.

The remaining chapters of the thesis are organized as follows. 
In chapter 2, we cover essential background knowledge and introduce another Monte Carlo method for rare-event estimation, Markov Chain Importance Sampling (MCIS).
We also present a survey of other common rare-event probability estimation techniques.  
In chapter 3, we draw analogies to the maximum likelihood estimation method to motivate the key mechanism underlying the ELM approach. 
This will be followed up by the formulation of the ELM procedure.
Further, in chapter 4, we perform numerical experiments using ELM and benchmark against existing robust estimators. 
In addition, critical analysis of strengths and shortcomings of the ELM approach is given.
Finally, in chapter 5, we provide concluding remarks and directions for future research. 
To preserve the flow of the thesis, we delegate various miscellaneous concepts and proofs to the appendix.

\chapter{Background}\label{background}

\section{Efficiency}

In this section, we explore the notion of efficiency in the context of rare-event probability estimation. This is important as we need a criteria to benchmark the performance of estimators.
A more efficient estimator will require smaller sample size to achieve a given degree of performance or accuracy. Firstly, the concept of {relative error} is introduced which we extend by incorporating simulation time. This will allow us to establish and compare efficiency of estimators in typical rare-event environments.

In statistical inference, efficiency of estimators is often characterized by its variance or mean square error which are key quantities in establishing the accuracy of the estimator. 
However, in a rare-event context, both the magnitude of variances and estimates are typically very small. 
Hence, we need to define a similar measure which captures the variability relative to the magnitude of the quantity we are trying to estimate. 
In other words, we need to borrow the idea of relative accuracy to give an indication of the estimator's performance relative to its magnitude. 
There exist several definitions in the literature \cite{asmussen2000rare,l2010asymptotic}. One measure we will be considering is the {relative error} defined below.

\paragraph{Relative Error:}
The relative error of an estimator $\widehat{\ell}$ is defined as its standard deviation divided by the quantity of interest $\ell$.
To formulate this, suppose $\widehat{\ell}$ is an unbiased estimator of the rare-event probability 
\begin{equation}\label{eq:rare_event_probability}
\ell=\mathbb{P}_f( S(\bX) \geq \gamma )=\Em_f \I\{ S(\bX) \geq \gamma \}, 
\end{equation}
where $f$ is a density function, $S$ is an arbitrary real-valued function, $\bX$ is a random vector and $\gamma$ is a level or threshold parameter. The quantity of interest is the expectation of the event $\{ S(\bX) \geq \gamma \}$ occurring under the density $f$.
Thus, the \emph{relative error} of the estimator $\widehat{\ell}$ is 
\begin{equation}
\RE(\widehat{\ell})=\frac{\sqrt{\Var(\widehat{\ell})}}{\ell}.
\end{equation}
The \emph{relative variance} or \emph{squared relative error} is 
\begin{equation}
\mathrm{RV}(\widehat{\ell})=\frac{{\Var(\widehat{\ell})}}{\ell^2}.
\end{equation}

To compute the above quantities,  we may need to estimate $\ell$ and $\Var(\widehat{\ell})$ by using realizations of the estimator.
Suppose we have $N$ independent replications, $Z_1,\dots,Z_N$, of some random variable $Z$ for which $\Em[Z]=\ell$.
Then an unbiased estimator of $\ell$ is
\begin{equation}
 \widehat{\ell}=\frac{1}{N}\sum_{i=1}^N Z_i.
\end{equation}
Further, note that $$\Var{(\widehat{\ell})}=\Var\left({\frac{1}{N}\sum_{i=1}^N Z_i}\right) = \frac{1}{{N}}\Var(Z).$$ Hence, the relative error of this estimator $\widehat{\ell}$ is 
\begin{equation} \label{RE_formula}
\RE(\widehat{\ell})=\frac{\sqrt{\Var(Z)}}{\ell \sqrt{N}}.
\end{equation}
This quantity can be estimated from the iid sample $Z_1, \dots, Z_N$ by replacing $\ell$ by its estimator $\widehat{\ell}$ and the variance by its sample counterpart
$$\widehat{{\Var(Z)}}={\frac{1}{N}\sum_{i=1}^N (Z_i-\widehat{\ell})^2}.$$
Similarly, the relative variance or squared relative error is
\begin{equation} \label{eq:relative_variance}
\mathrm{RV}(\widehat{\ell})=\frac{{\Var({Z})}}{N\ell^2}=\RE(\widehat{\ell})^2.
\end{equation}

It is important to note that the above efficiency measures ignore the simulation time, say $\tau$, to generate one realization of $Z$. Two estimators may have similar relative error but significantly different computational time. We can take into account the computational time by defining the {relative time variance product} below.

\paragraph{Relative Time Variance Product (RTVP):} 
In order to benchmark such estimators according to their efficiency, we must take into account simulation time along with its relative error.
Assume each of the $N$ replications takes equal amount of time $\tau$ to simulate. Then $N\tau$ is the total time to generate $N$ replications of $Z$.
Note $\tau$ is commonly measured in seconds. 
One approach to account for computational cost is to multiply the relative variance or squared relative error~\eqref{eq:relative_variance} by the total computational time $N\tau$.
It is given by
\begin{equation} \label{RTVP_formula}
\mathrm{RTVP}(\widehat{\ell})=N\tau \times \mathrm{RV}(\widehat{\ell})=\frac{\tau \Var(Z)}{\ell^2}.
\end{equation}
Hence, the relative time variance product of $\widehat{\ell}$  is independent of the number of replications $N$.
It is also known as the \emph{work normalized squared relative error}. 
This is an important measure that will  be used throughout our numerical experiments to compare estimators with differing computational costs and relative errors.

\section{Importance Sampling}

In typical rare-event settings, naive Monte Carlo methods are not viable due to inefficiency. A well known example is the Crude Monte Carlo (CMC) method where high computational costs are required to achieve a reasonable degree of relative error. This is illustrated in example~\reff{appendix:efficiency}{appendix:CMC} of the appendix. 
Efficiency, as measured by the size of RTVP, can be improved by either decreasing the simulation time to generate a replication or reducing the variance of the estimator. 
One popular approach for the latter is importance sampling. It is a widely used variance reduction technique in the estimation of \eqref{eq:Z_integral} and \eqref{eq:rare_probability_integral}.

Given a fixed degree of relative error, the variance reduction achieved through importance sampling is usually high enough such that its total computational effort is many orders of magnitude less than the CMC method. 
In many cases, the reduction in variance typically outweighs any increases in computational complexity yielding a smaller RTVP.

Importance sampling can be viewed as a special `acceleration' technique that allows us to change the probability measure. In effect, we can sample from another probability measure where rare-events occur more frequently. The procedure is described in detail below.

\subsection{Methodology}
Suppose $g$ is another probability density such that $Hf$ is dominated by $g$. In other words, for any $\bx$ such that $g({\bf x})  = 0$ then $H({\bf x})f({\bf x}) = 0$. 
We can rewrite \eqref{eq:Z_integral} as
$$ \mathcal{Z} = \int H({\bf x}) \frac{f({\bf x})}{g({\bf x})} g({\bf x}) \ d{\bf x} = \Em_g \left [ H({\bf X}) \frac{f({\bf X})}{g({\bf X})} \right ].$$
Thus, we take the expectation with respect to $g$ instead of $f$ and choose $g$ such that a reduction in variance of $\widehat{\mathcal{Z}}$ is achieved.  
Consequently, if $\bX_1, \bX_2,\dots, \bX_N$ is an iid population from $g$ then an unbiased estimator for $\mathcal{Z}$ is
\begin{equation}\label{eq:IS_estimator}
\widehat{\mathcal{Z}}=\frac{1}{N}\sum_{k=1}^N Z_k \quad \text{with} \quad Z_k = H({\bf X}_k)\frac{f({\bf X}_k)}{g({\bf X}_k)}.
\end{equation}
We call estimator $\widehat{\mathcal{Z}}$ the \emph{importance sampling estimator} and $g$ the \emph{importance sampling density}. The ratio of densities $$W({\bf x}) = \frac{f({\bf x})}{g({\bf x})}$$ is the \emph{likelihood ratio}.
If $\widehat{\sigma}$ is the sample standard deviation of $Z_1,\dots,Z_N$, then the relative error of $\widehat{\mathcal{Z}}$ is 
$$RE(\widehat{\mathcal{Z}}) = \frac{\widehat{\sigma}}{\sqrt{N}|\widehat{\mathcal{Z}}|},$$ and an asymptotic normal approximation $1-\alpha$ confidence interval for $\mathcal{Z}$ is 
$$\left (\widehat{\mathcal{Z}} - z_{1-\alpha/2}\frac{\widehat{\sigma}}{\sqrt{N}}, \widehat{\mathcal{Z}} + z_{1-\alpha/2}\frac{\widehat{\sigma}}{\sqrt{N}}\right), $$
where $z_{\omega}$ is the $\omega$--quantile of the standard normal distribution. 

The main difficulty in importance sampling is choosing an appropriate importance sampling density $g$ which yields an estimator with small variance. 
A poor choice of $g$ may compromise the quality of the estimate and its confidence intervals, see~\cite{asmussen2007stochastic}. 
This motivates the need for an optimal importance sampling density which minimizes the variance of the estimator in~\eqref{eq:IS_estimator}. Such importance sampling density exists in theory and is given in the next subsection below. 
We also note here that a good choice of $g$ should give the estimator~\eqref{eq:IS_estimator} finite variance. 
That is, $$\Em_g{H^2(\bX)\frac{f^2(\bX)}{g^2(\bX)}}=\Em_f{H^2(\bX)\frac{f(\bX)}{g(\bX)}}<\infty.$$
If possible, $g$ should be chosen such that it does not have lighter tails than $f$ and the likelihood ratio $f/g$ is bounded.

\subsection{Minimum-Variance Density}
The optimal importance sampling density $\pi$ is one that should minimize the variance of $\widehat{\mathcal{Z}}$.   
Thus, it is the solution to the functional minimization program
\begin{equation} 
\label{eq:IS_minimzation_program} 
\min_g{\text{Var}_g \left (  H({\bf X}) \frac{f({\bf X})}{g({\bf X})}\right) } 
\end{equation}
It is well known and can be shown, for example in~\cite{rubinstein1998modern}), that the solution to this optimization program is the \emph{minimum-variance importance sampling density}
\begin{equation} \label{eq:optimal_IS_pdf}
\pi({\bf x}) = \frac{| H({\bf x}) |f({\bf x})}{\int | H({\bf x}) |f({\bf x}) \ d {\bf x}}.
\end{equation}
In particular, if  $H({\bf x}) \geq 0 $ or $H({\bf x})  \leq 0$ for all $\bx$ then
\begin{equation}
\label{eq:zero_variance_density}
\pi({\bf x}) = \frac{H({\bf x}) f({\bf x}) }{\mathcal{Z}}
\end{equation}
is the \emph{zero-variance importance sampling density}. That is, our estimator $\widehat{\mathcal{Z}}$ is constant under $\pi$ and  in this case 
\begin{equation*}  
\text{Var}_{\pi}(\widehat{\mathcal{Z}}) = \text{Var}_{\pi}(H(\bX)W(\bX))=0.
\end{equation*}

\begin{Example}
In the context of rare-event estimation, we have assumed that $$H({\bf x})=\I\{S(\bx)\geq\gamma\}$$ which is an indicator function that maps to the set $\{0,1\}$. Thus, the property $H(\bx)\geq 0$ for all $\bx$ is satisfied. We can write the minimum (zero) variance importance sampling pdf in the form of~\eqref{eq:zero_variance_density}. That is, 
\begin{equation} \label{eq:min_var_IS}
\pi(\bx)=\frac{f(\bx) \I\{ S(\bx) \geq \gamma \} }{\ell},
\end{equation}
with $\ell = \mathbb{P}_f(S(\bX) \geq \gamma).$
\end{Example}

We note that in both~\eqref{eq:optimal_IS_pdf} and~\eqref{eq:zero_variance_density}, the optimal density $\pi(\bx)$ depends on the unknown quantity $\mathcal{Z}$ as $$\int | H({\bf x}) |f({\bf x}) \ d {\bf x}=\mathcal{Z}-2\int_{H(\bx)<0}H(\bx)f(\bx) \di\bx. $$ 
Hence, in practice, the evaluation of the minimum-variance importance sampling density $\pi$ is usually not possible and cannot be directly used as an importance sampling density. 
However, the minimum-variance density will play a significant role during the formulation of the ELM procedure in chapter~\ref{chapter3}.  

A good importance sampling density $g$ should still be `close' to the minimum variance density $\pi$. This motivates the concept of the  Kullback-Leibler distance given below.


\subsection{Kullback-Leibler Distance}
The optimal importance sampling density $\pi$ in~\eqref{eq:optimal_IS_pdf} or~\eqref{eq:zero_variance_density} is difficult to evaluate and cannot be used directly. Instead, consider an importance sampling density $g$ that is `close' to $\pi$. The closeness between two probability density functions $f$ and $g$ is commonly measured by the \emph{Kullback-Leibler Cross-Entropy Distance} which takes the form
\begin{equation}\label{eq:KLCE}
  \begin{split}
d(f,g) & = \Em \left [  \log \left(\frac{f(\bX)}{g(\bX)}\right)\right ] \\
       & = \int f(\bx) \log \left (\frac{f(\bx)}{g(\bx)} \right )  \di \bx \\
    		 & = \int f(\bx) \log \left (f(\bx) \right ) \ d\bx - \int f(\bx) \log \left (g(\bx) \right )  \di \bx.    
	\end{split}
\end{equation}
The Kullback Leibler distance is a special case of the $\phi$-divergence distance with $$\phi(x)=-\log(x).$$ This is explained further in appendix~\ref{appendix:distance_measures}, with other special cases of the $\phi$-divergence distance.

In theory, we would like to minimize the distance between the the minimum-variance density $\pi$ and our importance sampling estimator $g$
\begin{equation}\label{eq:distance_pi_g}
d(\pi,g)= \int \pi(\by) \log \left (\frac{\pi(\by)}{g(\by)} \right )  \di \by.
\end{equation}
However, without any restrictions, the optimal solution is $g(\by)=\pi(\by)$ which again is not useful. Instead, restrict the search space to product form 
$$g(\by)=\prod_{i=1}^d g_i(y_i).$$
Given this condition, it can be shown (see Lemma~\reff{MCIS}{lemma:POM}) that the best importance sampling density (in cross-entropy sense) is the product of the marginal densities from the minimum variance density
$$ g(\by)=\prod_{i=1}^d \pi_i(y_i).$$
This is inspired by the mean field approximation method in physics~\cite{kadanoff2009more} and is the backbone of the MCIS method discussed in the next section.

\section{MCIS Estimator}\label{MCIS}
In this section, we introduce an adaptive importance sampling procedure for rare-event probability estimation. The estimator combines two distinct and widely used Monte Carlo simulation methods, Markov chain Monte Carlo (MCMC) and importance sampling (IS), into a single algorithm. This is discussed in detail in~\cite{botev2011markov}.
The main steps of this algorithm are summarized as follows
\paragraph{1. MC Stage:} Construct the semi-parametric importance sampling density using MCMC samples from the zero-variance density~\eqref{eq:min_var_IS}.
\paragraph{2. IS Stage:} Use the constructed importance sampling density to deliver an estimator for $\ell$ by~\eqref{eq:IS_estimator}.

\subsection{Product of Marginals Model}
Instead of minimizing the Kullback-Leibler CE distance~\eqref{eq:distance_pi_g} over a simple parametric family of densities, we minimize it over all densities of product form
$$ g(\by) = \prod_{i=1}^d g_i(y_i), \quad g_i \in \mathcal{G}=\left\{ g: \mathbb{R}\rightarrow [0,\infty) \ \bigg | \ \int g(y) \ dy = 1 \right\}.$$
That is, we solve the functional optimization program
\begin{equation}\label{eq:FOP}
  \begin{split}
\min_{\substack{g_i \in \mathcal{G} \\ i=1,\dots,d}} d(\pi,g)=\int \pi(\by)\log \left ( \frac{\pi(\by)}{\prod_{i=1}^d g_i(y_i)}\right ) \ d \by.
  \end{split}
\end{equation}

\begin{Lemma} \label{lemma:POM} Suppose that $\pi_i(y_i)$ is the marginal of the zero-variance density $\pi(\by)$. Then, the solution to functional optimization program~\eqref{eq:FOP} is 
$$g_i(y_i)=\pi_i(y_i) \quad \text{for all} \quad i=1,\dots,d.$$
That is, the best importance sampling density (in the cross-entropy sense) within the space of all product-form densities is the product of marginals for~\eqref{eq:zero_variance_density}.
This is given by
\begin{align}
\label{eq:POMs}
g(\by) = \prod_{i=1}^d\pi_i(y_i).
\end{align}
\end{Lemma}

\begin{proof}
The proof of this lemma is given in appendix~\ref{appendix:lemma_POM}.
\end{proof}

\subsection{MCIS Methodology}
Suppose we have generated a population from the zero-variance density using the MCMC sampler. That is,
$$ \bX_1,\dots,\bX_n \stackrel{approx}{\sim} \pi(\bx), \quad \bX_i=(X_{i1},\dots,X_{id}). $$
We would like to evaluate~\eqref{eq:POMs}. Since the marginals are typically not available in closed form, we use the Markov chain output to estimate each marginal density $\pi_i(y_i)$ by
\begin{equation}\label{eq:estimator_MMVD}
\widehat{\pi}_i(y_i)=\frac{1}{n}\sum_{k=1}^n \pi(y_i|\bX_{k,-i}),
\end{equation}
where
\begin{itemize}
\item the population $\bX_1,\dots,\bX_n$ from $\pi$ is obtained approximately by the Gibbs Sampler;
\item the vector $\bX_{k,-i}$ is the same vector as $\bX_k$ with the $i$-th element removed;
\item $\pi(x_i|\bX_{k,-i})$ is the the conditional density of $x_i$ given all other components of $\bX_k$. 
\end{itemize}
The estimator~\eqref{eq:estimator_MMVD} is motivated by the identity
\begin{align*}
\Em_{\pi}[\widehat{\pi}_i(y)] 
& = \frac{1}{n}\sum_{k=1}^n \Em_{\pi}[\pi(y|\bX_{k,-i})] = \Em_{\pi}[\pi(y|\bX_{-i})] \\
& = \Em_{\pi} [\pi(y| X_1, \dots , X_{i-1}, X_{i+1}, \dots,X_d)] \\
& = \int \pi(y|x_1,\dots,x_{i-1},x_{i+1},\dots,x_d) \pi(\bx)  \di \bx \\
& = \int \frac{\pi(x_1,\dots,x_{i-1},y,x_{i+1},\dots,x_d)}{\pi(x_1,\dots,x_{i-1},x_{i+1},\dots,x_d)} \pi(\bx) \di \bx\\
& = \int \frac{\pi(x_1,\dots,x_{i-1},y,x_{i+1},\dots,x_d)}{\pi(x_1,\dots,x_{i-1},x_{i+1},\dots,x_d)} \left ( \int \pi(x_1,\dots,x_d) dx_i \right )  \di \bx_{-i}\\
& =  \int \pi(x_1,\dots,x_{i-1},y,x_{i+1},\dots,x_d) \di \bx_{-i} \\
& =\pi_i(y).
\end{align*}
We assumed that the conditional densities $\pi(x_i|\bX_{-i})$ are available in closed form. Thus, we can use the Gibbs sampling algorithm (see appendix~\ref{appendix_gibbs}) to obtain $\bX_1,\dots,\bX_n \stackrel{approx}{\sim} \pi(\bx)$.
Thus, in the MC stage we can construct the semi-parametric importance sampling density 
\begin{align}
\widehat{g}(\by) = \prod_{i=1}^d\widehat{\pi}_i(y_i) = \prod_{i=1}^d \frac{1}{n} \sum_{k=1}^n \pi(y_i|\bX_{k,-i}).
\end{align}
Generating $\bY \sim \widehat{g}(\by)$ is straightforward as each $Y_i$ is generated from the mixture $\widehat{\pi}_i(y_i)=\frac{1}{n} \sum_{k=1}^n \pi(y_i|\bX_{k,-i})$ and independently from other components of $\bY$ given $\{\bX_k\}$.
In the IS stage, we generate from the iid population $\bY_1,\dots,\bY_m$ from $\widehat{g}$ and then deliver the estimator 
\begin{align}\label{eq:MCIS_estimator}
\widehat{\ell}_{\text{MCIS}}=\frac{1}{m}\sum_{k=1}^m \I\{S(\bY_k)\geq\gamma\} \frac{f(\bY_k)}{\widehat{g}(\bY_k)}.
\end{align} 
Note here that the MCIS estimator uses the MCMC samples from the zero-variance density indirectly by first estimating $g$ and then $\ell$.
The estimator $\widehat{g}$ is semi-parametric as opposed to non-parametric since $\widehat{g}$ does not converge to $\pi$ as $n \uparrow \infty$ unless  the true $\pi$ is in product form. 


\section{Survey of Other Monte Carlo Techniques} \label{sec:survey}

A survey of rare-event simulation techniques and recent advances is given in~\cite{asmussen2007stochastic,juneja2006rare,kroese2011handbook}. 
In this section, we give a brief overview of the other common Monte Carlo methods used in rare-event probability estimation. 
This includes the adaptive importance sampling algorithms, conditioning methods for heavy tails and the splitting method.

\subsection{Adaptive importance sampling schemes} 
These schemes allow for the adaptive or automatic selection of the importance sampling density. One example is the MCIS procedure described previously in section~\ref{MCIS}.
Other examples include the cross-entropy and variance minimization methods, see~\cite{kroese2010cross,kroese2011handbook}.
We describe the well known cross-entropy method in more detail below.

\paragraph{Parametric Cross-Entropy Method:}
The key idea behind the cross-entropy method is to choose the importance sampling density $g$ in a specified parametric class of densities $\mathcal{G}$ 
such that the Kullback-Leiber divergence~\eqref{eq:KLCE} between the optimal importance sampling density $\pi$~\eqref{eq:optimal_IS_pdf} and $g$ is minimal.
This is described in detail in~\cite{kroese2010cross}.
We want to find a $g\in\mathcal{G}$ that minimizes
\begin{equation}
d(\pi,g)=\Em_{\pi} \left [ \log\frac{\pi(\bX)}{g(\bX)}\right ]= \int \pi(\by) \log \left (\frac{\pi(\by)}{g(\by)} \right ) \di \by.
\end{equation}
In most cases of interest the nominal probability density $f$ is parametrized by a finite-dimensional vector $\mathbf{u}$; 
that is, $f(\bx;\mathbf{u})$. It is then customary to choose the importance sampling density $g$ in the same family of probability densities. In other words, we would like 
$g(\bx) = f(\bx;\mathbf{v})$ for some reference parameter $\mathbf{v}$. The cross-entropy minimization procedure then reduces to finding an optimal reference parameter $\mathbf{v}^*$:
\begin{equation} \label{eq:ce_max}
\mathbf{v}^* =\argmin_{\mathbf{v}} \int \pi(\bx)\log\left(\frac{\pi(\bx)}{f(\bx;\mathbf{v})}\right)  \di \bx = \argmax_{\mathbf{v}} \int \pi(\bx) \log f(\bx;\mathbf{v}) \di\bx.
\end{equation}
In practice, the integral in~\eqref{eq:ce_max} is estimated from preliminary simulations so that we can obtain the estimator $\widehat{\mathbf{v}}^*$
\begin{equation}
\widehat{\mathbf{v}}^* =\argmax_{\mathbf{v}}\sum_{i=1}^n \log (f(\bX_i,\mathbf{v})),
\end{equation}
where $\bX_1, \dots, \bX_n$ is an approximate sample from the zero-variance density $\pi$~\eqref{eq:min_var_IS} obtained using MCMC sampling. For example, we can use the Gibbs Sampler; see algorithm~\reff{sec:MCMC}{algo:Gibbs} from the appendix.
Hence, we can deliver the cross-entropy estimator of~\eqref{eq:rare_probability_integral} as
\begin{equation}
\widehat{\ell}_{\text{CE}}=\frac{1}{m} \sum_{k=1}^m \I\{S(\bY_k)\geq\gamma\} \frac{ f(\bY_k;\mathbf{u})}{f(\bY_k;\widehat{\mathbf{v}}^*)}, \qquad \bY_1,\dots,\bY_m \stackrel{\text{iid}}{\sim} f(\by;\widehat{\mathbf{v}}^*).
\end{equation}

\subsection{Conditioning methods for heavy tails} \label{conditioning_method}
These are specialized and highly efficient algorithms designed for the estimation of probabilities arising from heavy-tailed random variables.
For characteristics of heavy-tailed distributions, see appendix~\ref{appendix:tail_properties}.
We focus on the Asmussen-Kroese conditional estimator~\cite[\S 10.3]{kroese2011handbook} below.
This is one of the estimators that will be used during the numerical benchmark problem in section~\ref{problem_statement}. 

\paragraph{Asmussen-Kroese (AK) Conditional Estimator:}
\label{AKCE}
This is an efficient estimator of rare-event probabilities that performs well for distributions with subexponential properties. 
It caters for estimation problems of the form $$ \ell(\gamma) = \PP(S_d{(\bX)} \geq \gamma) = \PP(X_1 + \cdots + X_d \geq \gamma),$$
where the $\{X_i\}$ are iid with cdf $F$ from the subexponential class of distributions. 
We are interested in the case where the threshold $\gamma$ is large, making  probability $\ell(\gamma)$ small. 
This estimator is based on a conditioning idea which exploits the subexponential property
$$\lim_{\gamma \rightarrow \infty} \frac{\PP{(S_d \geq \gamma)}}{d \ \PP{(X_1\geq \gamma)}}=1.$$
Typically, with the subexponential property, the rare-event primarily occurs due to a single variable exceeding the threshold value.
This is in contrast with the light-tailed case where the rare-event occurs usually when the majority of variables take on a large value.
The algorithm given below outputs the estimator and is based on the identity
$$\ell(\gamma) = d\ \PP(S_d(\bX)\geq \gamma, X_d = \max_{j}X_j) = d \  \Em\left [ \bar {F}  \left ( \left ( \gamma - \sum_{j=1}^{d-1} X_j \right )  \vee \max_{j\neq d } X_j\right ) \right ],$$
where $\bar{F}(x) = 1 - F(x)$ and $a\vee b  = \max\{a,b\}$. 
This is explained in more detail in~\cite{asmussen2006improved}.

\Algorithm{(AK Conditional Estimator for $\PP(S_d{(\bX)} \geq \gamma)$)}
\begin{enumerate}
\item Generate $X_1,\dots,X_n \stackrel{iid}{\sim} F.$
\item Compute $$Y = d \ \bar{F}\left (\left (\gamma - \sum_{j=1}^{d-1}X_j \right ) \vee \max_{j\neq d} X_j \right ) .$$
\item Using $N$ independent replications of $Y$, deliver the unbiased estimator 
\begin{equation}\label{eq:AK_estimator}
\widehat{\ell}(\gamma) = \frac{1}{N}\sum_{k=1}^N Y_k.
\end{equation}
\end{enumerate}
Hence, the AK conditional estimator based on one replication is given by
$$ \widehat{\ell}_{\text{AK}} = d \overline{F} \left ( \left (\gamma -\sum_{j=1}^{d-1}X_j  \right ) \vee \max_{j<d} X_j \right), \qquad X_1,\dots,X_{d-1} \stackrel{\text{iid}}{\sim} F.$$
We are interested in the Weibull distribution with scale parameter $1$ and shape parameter $\alpha<1$. This will be used in the numerical example later in section~\ref{problem_statement}.
In this case, the estimator~\eqref{eq:AK_estimator} has vanishing relative error (see  appendix~\ref{appendix:efficiency} for definition of vanishing relative error) when $0 < \alpha < \log(3/2)/\log(3)\approx 0.369$. For more detail, see~\cite[\S 10.3]{kroese2011handbook}.

\subsection{Splitting}
\label{splitting}

The splitting method, sometimes recognized as \emph{Sequential Monte Carlo (SMC)}, is another  simulation technique for the estimation of rare-event probabilities \cite{garvels2000splitting,glasserman1999multilevel}.
The main objective is to generate more occurrences of the rare-event by splitting the sample paths of (a possibly artificially induced) Markov chain into multiple copies at various stages of the simulation.
This method uses a decomposition of the state space into nested subsets so that the rare event is represented as the intersection of a nested sequence of events.
The probability of the rare event is the product of conditional probabilities,  each of which can be estimated much more accurately than the rare-event probability itself.
These methods are less effective than the conditioning methods used in section~\ref{conditioning_method} so we do not discuss them further. We note, however, that it may be possible to use the splitting method to generate approximate samples from $\pi$ as an alternative to MCMC.

\chapter{Theory of Empirical Likelihood Maximization}\label{chapter3}

In the MCIS algorithm, we have used the MCMC samples from the zero-variance density $\pi$ in \eqref{eq:min_var_IS} indirectly to first estimate the importance sampling density $g$ and then the (rare-event) probability $\ell$. 
While the estimator~\eqref{eq:MCIS_estimator} works well on a number of prototypical problems, we strive to do better by using the MCMC samples directly. This motivates a novel Monte Carlo technique -- \emph{Empirical Likelihood Maximization (ELM)}.
It is a versatile approach to rare-event probability estimation where we sample from a sequence of densities and maximize an empirical likelihood.
This method uses the MCMC samples directly. It is inspired by Bayesian methodology, where the objective is to estimate normalizing constants of posterior densities via MCMC sampling.


We begin by drawing some analogies to the maximum likelihood estimation method used ubiquitously in statistics.
The aim is to explain and motivate the key mechanism  underlying the ELM approach with the familiar maximum likelihood principle.
\section{Method of Maximum Likelihood}
Suppose we are given iid observed vectors $\bx_1,\dots \bx_n$ which we believe come from an unknown probability density function $g_0(\cdot)$.
However, we also believe that $g_0(\cdot)$ belongs to a certain parametric family of distributions $\{g(\cdot|\boldsymbol\theta):\boldsymbol\theta \in \boldsymbol\Theta\}$ such that 
$$g_0(\cdot) = g_0(\cdot|\boldsymbol\theta_0),$$
for some unknown value $\boldsymbol\theta_0$. 
We wish to estimate the true unknown $\boldsymbol{\theta_0}$. 
A desirable estimator $\boldsymbol{\widehat{\theta}}$ is one that is close in some sense to the true value $\boldsymbol\theta_0$. 
One way to obtain such an estimator is to use the \emph{maximum likelihood method}. 

To proceed, we specify the joint probability density function for the observed vectors. For an iid sample, this is given by
$$g(\bx_1,\dots,\bx_n | \boldsymbol\theta) = \prod_{i=1}^n g(\bx_i |\boldsymbol\theta).$$ 
Consider a different perspective where we hold the observed values $\bx_1,\dots \bx_n$ fixed, and allow $\boldsymbol \theta$ to vary.
This is known as the \emph{likelihood function} and is given by
$$L(\boldsymbol\theta|\bx_1,\dots,\bx_n) = \prod_{i=1}^n g(\bx_i; \boldsymbol \theta).$$
The maximum likelihood method searches the parameter space $\{\boldsymbol\theta \in \boldsymbol\Theta\}$ to find a value $\boldsymbol{\widehat{\theta}}$ that maximizes 
$L(\boldsymbol\theta|\bx_1,\dots,\bx_n)$.
We refer to this value as the  \emph{maximum likelihood estimator (MLE)} of $\boldsymbol\theta$ if it exists.
In other words, we need to solve the following optimization program to obtain the MLE
$$ \boldsymbol{\widehat{\theta}} = \argmax_{\boldsymbol\theta}L(\boldsymbol\theta|\bx_1,\dots,\bx_n).$$
It is often more convenient to work with the logarithm of the likelihood function; referred to as the \emph{log-likelihood}. 
Since logarithm is a monotonically increasing function, the same MLE $\boldsymbol{\widehat{\theta}}$ can also be obtained by maximizing the log-likelihood function
$$ \boldsymbol{\widehat{\theta}} = \argmax_{\boldsymbol\theta} \sum_{i=1}^n \log g(\bx_i;\boldsymbol\theta).$$ 
The key statistical properties of the MLE are as follows.
\begin{itemize}
\item Consistency -- The MLE $\boldsymbol{\widehat{\theta}}$ is consistent. 
That is, the estimator $\boldsymbol{\widehat{\theta}}$ converges in probability to its true value as $n \rightarrow \infty$
$$\boldsymbol{\widehat{\theta}} \stackrel{\mathbb{P}}{\rightarrow} \boldsymbol{\theta_0}.$$

\item Asymptotic Normality -- Under some regularity conditions, the MLE  $\boldsymbol{\widehat{\theta}}$  converges in distribution to a normally distributed random vector as $n\rightarrow\infty$
\begin{equation}
\label{MLE:Fisher}
\sqrt{n}(\boldsymbol{\widehat{\theta}}-\boldsymbol{\theta_0}) \stackrel{d}{\rightarrow} \textsf{N}(0,I^{-1}(\boldsymbol{\theta_0})), 
\end{equation}
where $I^{-1}(\boldsymbol{\theta_0})$ is the inverse of the Fisher information matrix.

\item Efficiency -- The MLE  $\boldsymbol{\widehat{\theta}}$ is asymptotically efficient as it attains the Cram\'er-Rao lower-bound
$$\lim_{n\rightarrow\infty}\Var{(\boldsymbol{\widehat{\theta}})}\ I(\boldsymbol{\theta})=1.$$

\end{itemize} 

\paragraph{Motivating the ELM approach:}

In the maximum likelihood estimation method, we found an estimator of the unknown parameter $\boldsymbol \theta$ by maximizing the likelihood function. 
The ELM procedure shares the same fundamental idea such that we maximize an (empirical) likelihood to obtain an estimator of the unknown (rare-event) probabilities of interest. 
One key difference is, instead of using a likelihood, we use an \emph{empirical likelihood} constructed from simulated data from known distributions or distributions known up to a multiplicative constant.

In the ELM method, we will use a sequence of probability densities with known or unknown normalizing constants. 
An important step is to embed the normalizing constants as parameters in the empirical likelihood.
These normalizing constants are treated as unknown parameters that are to be estimated. 
Hence, the empirical likelihood will be a function of embedded normalizing constants that do not affect the simulated observations.  
Maximizing the empirical likelihood or empirical log-likelihood with respect to such parameters will allow us to obtain an estimator of the (rare-event) probability.
That is, we want to solve the following optimization program
$$\widehat{\Bell} = \argmax_{\Bell} \left \{-\widehat{D}(\Bell;\bx)\right\},$$
where $-\widehat{D}(\Bell;\bx)$ is an empirical log-likelihood function and $\Bell$ is as vector of embedded parameters containing the quantity of interest.
Using knowledge of one known normalizing constant, we can retrieve estimates for the remaining parameters.
The resulting estimators would inherit consistency and asymptotic normality properties of the MLE.
More concrete and detailed mathematical explanations of the ELM procedure is given below.

\section{ELM Formulation} \label{sec:ELS_formulation}
Suppose we have a sequence of densities
\begin{equation}\label{eq:sequence_of_densities}
f_t(\bx) = \frac{w_t(\bx)}{\ell_t}=\frac{f(\bx)H_t(\bx)}{\ell_t}, \ t=1,\dots,s,
\end{equation}
where $f$ is a known density, $\{H_t\}$ are known functions and $\ell_t$ are normalizing constants to $\{w_t\}$.
A density  $f_t$ in the given sequence~\eqref{eq:sequence_of_densities} is a \emph{reference density} if it has a known normalizing constant that can be calculated analytically.
We assume the existence of at least one reference density $f_t$, say $f_1$, whose corresponding normalizing constant $\ell_1$ is known.

Now, we embed the rare-event probability of interest, say $\ell_s$, into this sequence. 
This can be achieved by using the zero-variance density~\eqref{eq:min_var_IS}
$$f_s(\bx)=\frac{f(\bx)\I\{S(\bx)\geq \gamma \}}{\ell_s}.$$
Hence, the problem of interest is to estimate the normalizing constant $$\ell_s = \Em_{f}[H_s(\bX)] = \mathbb{P}_{f}(S(\bX)\geq\gamma).$$
To proceed, we need to introduce a key condition required by the supports of the given sequence of densities.

\paragraph{Connectedness:} 
Suppose we are given a graph with $s$ nodes. Let there be an edge between two distinct nodes $i$ and $j$  if and only if 
\begin{align}
\label{eq:vardi_condition}
\Em_f \left [ \I\{H_i(\bX)>0\} \times \I\{H_j(\bX) > 0 \}  \right  ] > 0.
\end{align}
If there exists a path between any two nodes $i$ and $j$, then the condition~\eqref{eq:vardi_condition} is known as \emph{Vardi's connectivity condition}~\cite{gill1988large,vardi1985empirical} on the supports of $\{f_t\}$.
This connectivity condition is needed for justifying the results  presented in the subsequent sections.

\paragraph{Mixture Model:} 
Now, assume that iid samples from the sequence of densities $\{f_t \}$ have been simulated exactly, each with corresponding fixed sample size $n_t$. In other words, we have generated 
\[ 
\bX_{t,1},\dots,\bX_{t,n_t} \stackrel{iid}{\sim}f_t(\bx) \quad \text{for all} \quad t=1,\dots,s.
\]
Conceptually, this is similar to sampling $$ n = n_1 + \cdots + n_s $$ random variables with stratification from the mixture
\[
\label{eq:f_mix}
\bar{f}(\bx)=\frac{1}{n}\sum_{t=1}^s n_t f_t(\bx) = \sum_{t=1}^s \lambda_t f_t(\bx), \quad \lambda_t \stackrel{def}{=} \frac{n_t}{n}.
\]
Note $\lambda_t$ is fixed for each $t=1,\dots,s$ and is defined as the proportion of samples obtained from density $t$. 
Now, to group the observations together, let the pooled sample be $$\bX_1, \dots, \bX_n$$ where the first $n_1$ samples are realizations from $f_1$, the next $n_2$ are samples from $f_2$ and so on. 
In practice, sampling may need to be done approximately (e.g. via MCMC, see appendix~\ref{sec:MCMC}) if exact sampling methods (see appendix~\ref{random_variable_simulation}) are not viable.   
In the following step of the ELM procedure, we will use the samples to construct the empirical likelihood.

\paragraph{Empirical Likelihood:}
To proceed, we define the vector of parameters 
\[
\mathbf{z}=(z_1,\dots,z_n)=\left(-\log\left(\frac{\ell_1}{\lambda_1}\right),-\log\left(\frac{\ell_2}{\lambda_2}\right),\dots,-\log\left(\frac{\ell_s}{\lambda_s}\right)\right)=-\log\left(\frac{\boldsymbol \ell}{\boldsymbol \lambda}\right).
\]
Now, consider the empirical likelihood of the observed simulated data $\bX_{k,1},\dots,\bX_{k,n_k}$, $k=1,\dots,s$, as a function of the parameter vector $\mathbf{z}$
\begin{align} \label{complete_likelihood}
\widehat{L}(\mathbf{z}) & = \prod_{k=1}^s\prod_{j=1}^{n_k} f_k(\bX_{k,j}) \notag \\ 
& = \prod_{k=1}^s\prod_{j=1}^{n_k}\frac{w_k(\bX_{k,j})}{\lambda_k e^{-z_k}}, \quad \text{(as $\ell_k =\lambda_k e^{-z_k}$ )} \notag \\
& = \prod_{k=1}^s\prod_{j=1}^{n_k} \frac{w_k(\bX_{k,j})}{\lambda_k e^{-z_k} \bar{f}(\bX_{k,j})} \ \prod_{k=1}^s\prod_{j=1}^{n_k} \bar{f}(\bX_{k,j}).
\end{align}
Further, the partial empirical log-likelihood (by~\cite{gill1988large}) is
\begin{align} \label{eq:partial_empirical_LL}
\widehat{l}(\mathbf{z}) & =  \log \left ( \prod_{k=1}^s\prod_{j=1}^{n_k} \frac{w_k(\bX_{k,j})}{\lambda_k e^{-z_k} \bar{f}(\bX_{k,j})} \right ) \notag\\
& = \sum_{k=1}^s\sum_{j=1}^{n_k}\left ( \log w_k(\bX_{k,j})  + z_k - \log \lambda_k - \log \bar{f}(\bX_{k,j})\right) \notag\\
& = -\sum_{j=1}^n \log \bar{f}(\bX_{j}) + \sum_{k=1}^s n_k z_k + C \notag \\
& = -\sum_{j=1}^n \log \left (\sum_{k=1}^s w_k(\bX_j)e^{z_k} \right ) + \sum_{k=1}^s n_k z_k + C, 
\end{align}
where $$C=\sum_{k=1}^s \sum_{j=1}^{n_k} \log w_k(\bX_{k,j}) + \sum_{k=1}^s n_k \log \lambda_k$$ is a constant independent of the parameters $\mathbf{z}$.
Define
\begin{equation}\label{eq:objective_function}
\widehat{D}(\mathbf{z})\stackrel{def}{=}\sum_{j=1}^n \log \left (\sum_{k=1}^s w_k(\bX_j)e^{z_k} \right )  - \sum_{k=1}^s n_k z_k,
\end{equation}
which is a convex function of $\bz$. This will be our objective function in the optimization program formulated below.

\paragraph{Optimization:}
We observe that maximizing the partial log-likelihood in~\eqref{eq:partial_empirical_LL} is equivalent to minimizing~\eqref{eq:objective_function}.
Under Vardi's connectivity condition~\eqref{eq:vardi_condition} and assumption of an iid sample, it is shown~\cite{gill1988large,vardi1985empirical} that the maximum of this partial empirical log-likelihood is same as the maximum of the complete empirical likelihood~\eqref{complete_likelihood}. 

Also, under the same conditions, it is shown~\cite{gill1988large,vardi1985empirical} that the unique nonparametric likelihood estimate of $\widehat{\mathbf{z}}$ (and hence $\widehat{\Bell}$) is the solution to the convex optimization program
\begin{equation}\label{eq:unconstrained_optimization}
  \begin{split}
\mathbf{\widehat{z}} & = \argmin_{\mathbf{z}} \widehat{D}(\mathbf{z}), \\
  \end{split}
\end{equation}
with $\ell_1$ known and $z_1$ fixed. 
This restriction is required to obtain an unique and optimal solution $\widehat{\mathbf{z}}$ from the parameter space.
We will see later, in section~\ref{constrained_MEL}, that such restriction can be replaced by adding in a homogeneous constraint in the optimization.

Solving the unconstrained program~\eqref{eq:unconstrained_optimization} will allow us to obtain an optimal solution $\widehat{\mathbf{z}}$ and hence $\widehat{\Bell}$ up to a multiplicative constant. Having known $\ell_1$, we are able to recover the remaining $\widehat{\Bell}$ and hence $\widehat{\ell}_s$. The idea is similar to recovering an unique eigenvector by knowing its first component (an eigenvector corresponding to an eigenvalue of multiplicity unity is unique up to a multiplicative constant). 
One way to execute this convex optimization is to compute the gradient of the objective function and equate it to zero.
This leads to the moment-matching equations described in detail below.

\section{Moment-Matching Equations} \label{section:moment_matching}
Without loss of generality, assume the known normalizing constant $\ell_1$ of the reference density $f_1$ is equal to unity. 
Then, the gradient is a $(s-1)\times1$ vector with components
\begin{equation} 
\begin{split} 
[\nabla \widehat{D}]_t (\mathbf{z})= \sum_{j=1}^n \frac{w_t(\bX_j) e^{\widehat{z}_t}}{\sum_{k=1}^s w_k(\bX_j) e^{\widehat{z}_k}} - n_t,   \quad \text{for} \quad t = 2,\dots,s.
\end{split}
\end{equation}
Hence, we equate these to zero and solve the $s-1$ dimensional system for the unknowns $z_2,\dots,z_s$
\[
[\nabla \widehat{D}]_t (\log(\lambda_1), z_2, \dots, z_s) =0, \quad \text{for} \quad t = 2,\dots,s.
\]
In other words, solve the following nonlinear system for   $\widehat{z}_2,\dots,\widehat{z}_s$
\begin{equation}
\frac{1}{n} \sum_{j=1}^n \frac{w_t(\bX_j) e^{\widehat{z}_t}}{\sum_{k=1}^s w_k(\bX_j) e^{\widehat{z}_k}} = \lambda_t, \qquad t = 2,\dots,s.
\end{equation}
The solution of these equations will include $\widehat{z}_s$ and hence this will give us an estimator of the rare-event probability of interest $\widehat{\ell}_s$.
Alternatively, the nonlinear system $[\nabla \widehat{D}]_t (\widehat{\mathbf{z}}) = 0$ for $t=2,\dots,s$ can be written in terms of the vector $\widehat{\Bell}$ 
\begin{equation} \label{eq:moment_matching_eqns}
\widehat{\ell}_t=\frac{1}{n}\sum_{j=1}^n \frac{w_t(\bX_j)}{\sum_{k=1}^s \lambda_k f_k(\bX_j)}=\frac{1}{n_t}\sum_{j=1}^n \frac{\lambda_t w_t(\bX_j)}{\sum_{k=1}^s \lambda_k f_k(\bX_j)}, 
\qquad t=2,\dots,s.
\end{equation}
This resembles the \emph{generalized method of moments (GMM)} (see~\cite{hansen1982large}) for estimating parameters in statistical models. 
A brief description is given below.

\paragraph{Generalized Method of Moments:}
Suppose we have $n$ iid observations $\bX_1,\dots,\bX_n$ where each observation is an $d$-dimensional multivariate random variable. 
We assume the data comes from a certain statistical model with an unknown parameter $\boldsymbol{\theta}\in\boldsymbol{\Theta}$. 
The aim of the estimation problem is to find the `true' value of this parameter, $\boldsymbol\theta_0$.
In order to apply GMM there should exist a vector-valued function $\mathbf{g}(\bX,\boldsymbol\theta)$ such that
$$ \mathbf{m}(\boldsymbol\theta_0)\stackrel{def}{=}\Em_{\boldsymbol\theta_0}[\mathbf{g}(\bX;\boldsymbol\theta_0)] =\mathbf{ 0}. $$
The central idea behind GMM is to replace the theoretical expected value with its empirical first moment (the sample average)
$$\widehat{\mathbf{m}}(\boldsymbol\theta)=\frac{1}{n}\sum_{i=1}^n \mathbf{g}(\bX_i;\boldsymbol\theta),$$
and then to minimize the norm of this expression with respect to $\boldsymbol\theta$.
Similarly, the identity~\eqref{eq:moment_matching_eqns} can be rewritten such that it resembles the matching of $s$ empirical first moments.
In other words, we solve the following system of non-linear equations
$$ \widehat{\mathbf{m}}_{t-1}(\boldsymbol\ell)\stackrel{def}{=}\frac{1}{n_t}\sum_{j=1}^n \frac{\lambda_t w_t(\bX_j)}{\sum_{k=1}^s \lambda_k f_k(\bX_j)} - {\ell}_t=0, \qquad  t=2,\dots,s. $$ 
However, instead of minimizing the norm, this nonlinear system can be solved exactly by using Jacobi or Gauss-Seidel iterations given below in algorithm~\reff{sec:ELS_formulation}{algo:Jacobi_iterations}.
Executing this will allow us to obtain $\widehat{\Bell}$ up to a multiplicative constant. 
In effect, using the knowledge of $\ell_1$, we can determine the remaining $\widehat{\ell}_t$ for $t=2,\dots,s$.
This process is similar to finding eigenvalues via power iterations. 

\begin{Algorithm} (Jacobi Iterations) \label{algo:Jacobi_iterations}\\
Require: Matrix $A$ and initial starting point $\Bell=(\ell_1,\dots,\ell_s)=(1,\dots,1)$ \\
Set $\epsilon = \infty$ and $\Bell^* = \Bell$ \\
While $\epsilon >$  tol do \\
\indent For $i=2,\dots,s$ do \\
\indent \indent $\ell_i = \sum_{j=1}^n \frac{A_{i,j}}{\sum_{k=1}^s A_{k,j} n_k / \ell^*_k},$  \qquad $A_{i,j} \stackrel{def}{=} H_i(\bX_j). $\\
\indent End do \\
\indent $\epsilon = \max_i\frac{\ell_i - \ell^*}{\ell_i} $ \\
\indent Set $\Bell^* = \Bell$ \\
End do\\
Return the vector of estimated probabilities $\widehat{\Bell} = \Bell$.
\end{Algorithm}

Although  the Jacobi algorithm above works satisfactorily, it sometimes takes many iterations to converge. 
This is why, in subsequent sections, we offer an faster alternative to minimizing $\widehat D$ ---  direct optimization of $\widehat D$ using {\tt MATLAB}'s {\tt fmincon}  routine. 

\section{Constrained Maximum Empirical Likelihood}
\label{constrained_MEL}
Instead of solving the moment-matching equations to obtain $\widehat{\ell}$, it is sometimes easier to use an optimization routine to minimize $\widehat{D}$ directly.  
In effect, we are able to add in constraints along with the optimization. 
This is a powerful approach to incorporate any analytical knowledge for the given sequence of densities $\{ f_t, t=1,\dots,s \}$ which is one of the key strengths of the ELM method.
We begin by explaining how equality constraints can be added, followed by inequality constraints. 

\paragraph{Equality Constraints:}

Recall that we required the first component of $\bz$, $z_1$, to be fixed in the optimization program~\eqref{eq:unconstrained_optimization}. 
This is needed to find an unique solution $\widehat{\mathbf{z}}$ in the parameter space.
Alternatively, we can add in a \emph{homogeneous constraint} to alleviate this restriction
\begin{equation} \label{homogeneous_constraint}
\lambda_1z_1+\lambda_2z_2 + \cdots + \lambda_s z_s = 0.
\end{equation}
This linear constraint is \emph{homogeneous} in the sense that the linear combination of $z_1,\dots,z_s$ is equal to zero.  
Solving the optimization problem using only this homogeneous constraint is equivalent to solving the Jacobi iterations in~\reff{section:moment_matching}{algo:Jacobi_iterations}.

We can also add in an \emph{inhomogeneous} linear constraint where the linear combination of $z_1,\dots,z_s$ does not need to be zero. 
This will allow us to incorporate any analytical knowledge of reference densities with known normalizing constants.
For example, suppose that two of the normalizing constraints, say $\ell_1$ and $\ell_2$, are known analytically.
Then we can add in the following equality constraint
$$z_1 - z_2 = \log(\lambda_1) - \log(\lambda_2)+\log(\ell_2)-\log(\ell_1).$$
This will allow us to use $\ell_1$ and $\ell_2$ to reduce variance and improve the quality of the estimator.
Hence, we can solve the following constrained optimization to estimate the remaining $\ell_t, t=3,\dots,s$
\begin{equation}\label{eq:constrained_program}
\begin{split}
& \min_{\mathbf{z}} \widehat{D}(\mathbf{z}) \\
& \text{subject to:} \\
& \lambda_1z_1+ \cdots + \lambda_s z_s = 0, \\ 
&z_1 - z_2 = \log(\lambda_1) - \log(\lambda_2)+\log(\ell_2)-\log(\ell_1).
\end{split}
\end{equation}
Solving the constrained program~\eqref{eq:constrained_program} will allow us to obtain an optimal solution $\widehat{\mathbf{z}}$ and hence $\widehat{\Bell}$ up to a multiplicative constant.
Using the analytical knowledge of known $\ell_1$ or $\ell_2$, we are able to recover the remaining $\widehat{\Bell}$ and hence $\widehat{\ell}_s$. 
It is also easy to extend this idea by adding in more equality constraints which would allow for additional idiosyncratic knowledge of relationships amongst reference densities.

\paragraph{Inequality Constraints:}
Similar to the equality constraints, it is also simple to add inequality constraints.
This will allow us to include any analytical knowledge of lower or upper bounds to the probabilities that are to be estimated.
For example, suppose one of the reference densities, say $f_1$, has a normalizing constant $\ell_1$ which serves as a lower bound to $\ell_s$. That is,
$$\ell_1 \leq \ell_s.$$
To incorporate this, we just need to add the inequality constraint
$$-z_1 + z_2 \leq -\log(\lambda_1)+\log(\lambda_2).$$
Hence, along with the homogeneous constraint~\eqref{homogeneous_constraint}, our constrained optimization program becomes
\begin{equation}\label{eq:inequality_constrained_program}
\begin{split}
& \min_{\mathbf{z}} \widehat{D}(\mathbf{z}) \\
& \text{subject to:} \\
& \lambda_1z_1+ \cdots + \lambda_s z_s = 0, \\ 
&-z_1 + z_2 \leq -\log(\lambda_1)+\log(\lambda_2).
\end{split}
\end{equation}
Again, solving this constrained program delivers an optimal solution $\widehat{\mathbf{z}}$ and hence $\widehat{\Bell}$ up to a multiplicative constant. 
Hence, we are able to recover  $\widehat{\ell}_s$ by using analytical knowledge of known components of $\Bell$.

\subsection{Standard Errors and Asymptotic distribution}
It is possible to derive a central limit theorem for the estimator $\widehat \bz$ (which solves the constrained (empirical) likelihood problem \eqref{eq:constrained_program}), similar to the classical likelihood one given in \eqref{MLE:Fisher}. This will allow us  to derive asymptotic standard errors for $\widehat \bz$ using the corresponding Fisher information matrix. However, such limiting results will rely on the iid assumption on the sample $\bX_1,\ldots,\bX_n$ and we will be using an approximate dependent sample from MCMC. 
Of course, it is possible to construct central limit results in the MCMC case involving estimation of  an autocorrelation function,  but this is beyond the scope of an Honours thesis.  For this reason and to keep things simple, we leave such derivations as future research. In this thesis, we will estimate standard errors from repeated  simulation runs of the ELM algorithm providing us with a sequence of $\widehat \bz_1,\ldots, \widehat\bz_{K}$. 


\chapter{Application to Rare-event Probability Estimation}
In this chapter, we illustrate the effectiveness and versatility of the ELM approach by focusing on a numerical example: sum of heavy-tailed iid Weibull random variables. 
Computationally, this is of interest because there exist algorithms that efficiently estimate $\ell$ for certain values of $\alpha$ but not its entire range. 
For example, the Asmussen-Kroese conditional estimator described in section~\ref{sec:survey} has vanishing relative error property when $0 <\alpha<\log(3/2) / \log(3) \approx 0.369$, see appendix~\ref{appendix:efficiency}. It is known that this estimator is less efficient and can fail when $\alpha$ falls outside the given range. 
We will demonstrate that the ELM estimation method performs well for all values of $\alpha<1$. 
This is a numerical benchmark problem from~\cite{asmussen2006improved} and is stated below.

\section{Sum of heavy-tailed iid  Weibull random variables}
\label{problem_statement}
Suppose that $\bX = (X_1,\dots,X_d)$ is a $d$-dimensional independent and identically distributed random vector where each component is from the $\textsf{Weib}(\alpha,1)$ distribution for some fixed shape parameter $\alpha$. Denote the joint $\textsf{Weib}(\alpha,1)$ distribution by 
$$f(\bx) = \prod_{i=1}^d \alpha x_i^{\alpha-1}\e^{-x_i^{\alpha}}, \ x_i \geq 0 \quad \forall i\in\{1,\dots,d\} .$$ 
Each random component from this distribution is heavy-tailed when $\alpha<1$. It is of interest to estimate the (rare-event) probability 
\[ \ell = \Em_f \I(S(\bX) \geq \gamma) = \mathbb{P}(S(\bX) \geq \gamma), \]
with $$S(\bX) = \sum_{i=1}^d X_i.$$ For example, this may represent the probability of default for an insurance company where $S(\bX)$ is the sum of all claims received and $\gamma$ is their current level of capital.

We consider two different ELM schemes.
The first ELM scheme given in~\ref{ELS_Scheme1} below uses 4 densities and incorporates equality constraints.
The densities included are ones which we believe work best in obtaining an efficient ELM estimator. 
Further, the second ELM given in~\ref{ELS_Scheme2}  uses 2 densities and incorporates an inequality constraint.
We will present a novel approach in obtaining a lower bound to our (rare-event) probability of interest.
This is the underlying mechanism behind the second ELM scheme allowing us to utilize an inequality constraint.


\section{ELM Scheme A: 4 densities with equality constraints}
\label{ELS_Scheme1}
We apply the procedure as described in section~\ref{sec:ELS_formulation} and~\ref{constrained_MEL} with $s=4$ densities.

\subsection{ELM Densities}
The four ELM densities, to be introduced, are key ingredients in obtaining an efficient ELM estimator $\widehat{\ell}_s$. 
Firstly, we embed the (rare-event) probability of interest $\ell_s$ into $f_s$ by using the zero-variance density from~\ref{eq:min_var_IS}.
To motivate the inclusion of the other densities, consider rewriting the (rare-event) probability as 
$$\ell_s = \mathbb{P}(S(\bX) \geq \gamma) = \mathbb{P}(M(\bX) \geq \gamma) + \mathbb{P}(S(\bX) \geq \gamma, M(\bX) < \gamma),$$
where $$M(\bX) = \max_{i\in\{1,\dots,d\}} X_i=X_{[d]}.$$
This separates the (rare-event) probability into its main `heavy-tailed' component of $\ell_s$  
$$\mathbb{P}(M(\bX) \geq \gamma)$$ which represents the probability that the largest element of the random vector exceeds the $\gamma$ threshold, and its residual component of $\ell_s$ 
$$\mathbb{P}(S(\bX) \geq \gamma, M(\bX) < \gamma) $$ which represents the probability that the sum of the elements exceeds the $\gamma$ threshold but its largest component $X_{[d]}$ will not.
Generally, for heavy-tailed distributions, the magnitude of the main components greatly exceeds the residual component. That is,
$$\lim_{\gamma\uparrow\infty} \frac{\mathbb{P}(S(\bX) \geq \gamma, M(\bX) < \gamma)}{\mathbb{P}(M(\bX) \geq \gamma)} \rightarrow 0.$$
This is particularly the case when $\alpha$ is close to zero for this $\textsf{Weib}(\alpha,1)$ example. 
However, the residual component becomes more significant when the distribution becomes less heavy-tailed (i.e. when $\alpha$ is close to $1$).
The Asmussen-Kroese conditional estimator given in section~\ref{sec:survey} captures the main `heavy-tailed' contribution very well but ignores the residual component.


The first two densities are reference densities such that their normalizing constants $\ell_1$ and $\ell_2$ are known analytically. 
We can use such information as a correction factor during the constrained empirical likelihood optimization. 
The normalizing constant of the third density $\ell_3$ is unknown but the samples attained allow us to capture the residual component of the probability of interest. 
More detail of these four densities are given below.

\begin{enumerate}
\item Reference density: Main `heavy-tailed' component (sum of the indicators) $$f_1(\bx) = \frac{f(\bx) \sum_{i=1}^d \I \left \{ x_i \geq \gamma \right\}}{\ell_1}= \frac{w_1(\bx)}{\ell_1},$$
where $$\ell_1 =\sum_{i=1}^d \mathbb{P}(X_i \geq \gamma) = d\cdot \mathbb{P}\left( X \geq \gamma\right) = d(1-F_X(\gamma))=de^{-\gamma^\alpha},$$ 
with $X\sim\textsf{Weib}(\alpha,1).$ 
This density is introduced to capture the main contribution of $\ell_s$. It has normalizing constants that represent the sum of the probabilities for each component exceeding the $\gamma$ threshold.

\item Reference density: Product of marginals from the zero-variance density~\eqref{eq:min_var_IS} $$f_2(\bx) = \prod_{i=1}^d \widehat{\pi}_i(x_i)= \frac{w_2(\bx)}{\ell_2},$$
where $\widehat{\pi}_i(x_i)$ are the marginal densities of $f_s$. Note $\ell_2=1$.
This is the importance sampling density used in the MCIS method (see section~\ref{MCIS}). 

\item Density: Residual component
$$f_3(\bx)=\frac{f(\bx) \I\{S(\bx) \geq \gamma, M(\bx) < \gamma\}}{\ell_3}=\frac{w_3(\bx)}{\ell_3},$$
where $$M(\bx) = \max_{i\in\{1,\dots,d\}} x_i=x_{[d]}.$$ 
This density captures the residual component of $\ell_s$.
For each sample obtained from $f_3$, the sum of the elements will exceed the $\gamma$ threshold but its largest component $X_{[d]}$ will not.
Here, the normalizing constant $\ell_3$ is unknown and represents the residual (rare-event) probability.

\item Target density: minimum (zero) variance density from~\eqref{eq:min_var_IS}, $s=4$,
\[ f_s(\bx) = \frac{f(\bx) \I\left\{S(\bx) \geq \gamma \right\}}{\ell_s} = \frac{w_s(\bx)}{\ell_s},\]
where $$\ell_s=\Em_f [ \I(S(\bX) \geq \gamma) ] = \mathbb{P}(S(\bX) \geq \gamma)$$ is unknown and represents our (rare-event) probability of interest.
\end{enumerate}

\subsection{Sampling}
In this subsection, we describe the sampling procedures used to generate random realizations from each of our four densities. Denote ${n_{t}}$ to be the corresponding sample sizes for each density $t=1,\dots,4$ respectively and define our pooled sample to be $\bX_1,\dots,\bX_n$ where $n = n_1 + \dots+n_4$. 

\paragraph{Sampling from $f_1$:}
Sampling iid copies from the reference density $f_1$ is easy by using its mixture representation
$$f_1(\bx) = \sum_{j=1}^d \frac{\mathbb{P}(X_j \geq \gamma)}{\ell_1} \frac{f(\bx)\I\{ x_j \geq \gamma\}}{\mathbb{P}(X_j \geq \gamma)}, $$
where we select component $J$ with probability $\mathbb{P}(J=j) = \mathbb{P}(X_j \geq \gamma) / \ell_1$.
Given $J=j$, we sample from the truncated joint pdf $$ \frac{f(\bx | x_j) f(x_j)\I\{ x_j \geq \gamma\}}{\mathbb{P}(X_j \geq \gamma)}, $$
where each $X_i$ for $i\neq j$ has iid $\textsf{Weib}(\alpha,1)$ and $X_j$ has truncated pdf  
$$ \frac{f(x_j)\I\{ x_j \geq \gamma\}}{\mathbb{P}(X_j \geq \gamma)}, \quad 0 < \gamma \leq x_j < \infty. $$
The algorithm below describes how this can be implemented.

\begin{Algorithm}
\begin{enumerate}
\item Select $j \in \{1,\dots d\}$ with equal probability.
\item Generate $U \sim$ {\sf Uniform}($0,1$).
\item Set $X_j := (\gamma^\alpha - \ln U)^{\frac{1}{\alpha}}$ as our sample from truncated Weibull.
\item Set $X_i := (-\ln U)^{\frac{1}{\alpha}}$ for all $i \neq j$ as our sample from Weibull.
\item Return $ \bX = (X_1, \dots, X_d)$.
\end{enumerate}
\end{Algorithm}
Here, we obtain $n_1$ replications $\bX_{1}, \dots, \bX_{n_1}$ from $f_1$. 

\paragraph{Sampling from $f_2$:}
As $f_2$ is a product of independent marginals, we can sample each element of $\bX$ independently.
Note that $\widehat{\pi}_i(x_i)$ can be interpreted as a mixture of conditional densities using samples $\bX_j$ from $f_s$. 
Hence, it suffices to sample from one of $f_s(x_i | \bX_{j,-i})$, $j=1,\dots,n_s$ with equal probability. Each element of our sample will be from marginal or truncated $\textsf{Weib}(\alpha,1)$.
To attain a sample $\bX=(X_1,\dots,X_d)$, we utilize the following algorithm.

\begin{Algorithm} \\
For all $i = 1,\dots, d$, perform the following
\begin{enumerate}
	\item Generate $j \in \{1,\dots,n_s \}$ with equal probability.
	\item To sample from $f_s(x_i | \bX_{j,-i})$, compute $c_{ji} = \sum_{k\neq i} X_{jk}$.
	\item Generate $U \sim \textsf{Uniform}(0,1)$.
	\item Set $$ x_i := 
	\begin{cases}
	(-\log U)^{1/\alpha} & \ \text{if} \ c_{ji} > \gamma, \\
	\left [ (\gamma - c_{ji})^\alpha_{+} - \log U \right ]^{1/\alpha} & \ \text{if} \ c_{ji} \leq \gamma,
	\end{cases}$$
	where $(y)_{+} = \max\{0,y\}$.
\end{enumerate}
\end{Algorithm}
Here, we generate $n_2$ samples $\bX_1, \dots, \bX_{n_2}$ from $f_2$.

\paragraph{Sampling from $f_3$:}
We apply the Gibbs Sampler, see algorithm~\reff{sec:MCMC}{algo:Gibbs}. Let $\bX_{-i} =(X_1,\dots,X_{i-1},X_{i+1},\dots x_d)$ and $a_+:=\max\{0,a\}$.
The conditional densities for all components $i = 1,\dots, d$ are truncated $\textsf{Weib}(\alpha,1)$ given by
$$f_3(x_i | \bX_{-i}) \propto {f(x_i)}, \quad \left(\gamma - \sum_{j \neq i} X_j \right)_{+} \leq x_i < \gamma, $$
where $f(x_i)$ is the $\textsf{Weib}(\alpha,1)$ density.
To sample from the conditional densities, we can use the inverse-transform method for truncated distributions (see algorithm~\reff{random_variable_simulation}{algo:truncation}). 
For these truncated Weibull, our samples are obtained by setting for all $i = 1,\dots, d$
$$ X_i = \left [  -\log \left (e^{-a^{\alpha}} - U(e^{-a^{\alpha}}-e^{-\gamma^{\alpha}})  \right ) \right ]^{\frac{1}{\alpha}}, $$
where
$$a = \left(\gamma - \sum_{j \neq i} X_j \right)_{+}  $$
and $$U\sim \textsf{Uniform}(0,1).$$ Here, we generate $n_3$ samples $\bX_1, \dots, \bX_{n_3}$ from $f_3$.

\paragraph{Sampling from $f_s$:}
This is a special case of the above, where the conditional densities for all $i = 1,\dots, d$ are truncated $\textsf{Weib}(\alpha,1)$ of the form
$$f_s(x_i | \bX_{-i}) \propto {f(x_i)}, \quad \left(\gamma - \sum_{j \neq i} X_j \right)_{+} \leq x_i < \infty. $$
To sample from these conditional densities, set for all $i = 1,\dots, d$
$$ X_i = \left [  (\gamma - \sum_{j\neq i}{x_j} )_+^\alpha  - \ln U \right ]^{\frac{1}{\alpha}}. $$
Here, we generate $n_s$ samples $\bX_1, \dots, \bX_{n_s}$ from $f_s$.
Hence, we obtain the pooled sample $\bX_1,\dots,\bX_n$. 
We can now proceed to formulate the empirical likelihood and perform the optimization.


\subsection{Empirical Likelihood Optimization}
Suppose we have generated ${n_{t}}$ samples for each density $t=1,\dots,4$ and have obtained a pooled sample $\bX_1,\dots,\bX_n$ where $n = n_1 + \dots+n_4$. Define the vector of parameters
$$\mathbf{z} = (z_1,\dots,z_n)=(-\log(\ell_1/\lambda_1),\dots,-\log(\ell_s/\lambda_s)).$$ 
We want to solve the following constrained nonlinear convex optimization program
\begin{equation}
\begin{split}
& \min_{\mathbf{z}} \widehat{D}(\mathbf{z}) \\
& \text{subject to:} \\
& \sum_{i=1}^s \lambda_i z_i  = 0 \\ 
&z_1 - z_2 = E_{12}, 
\end{split}
\end{equation}
where $$ E_{12} = \log(\lambda_1) - \log(\lambda_2)+\log(\ell_2)-\log(\ell_1)=\log(\lambda_1) - \log(\lambda_2) - \log(d) + \gamma^{\alpha}. $$
The objective function is
\begin{align*}
\widehat{D}(\mathbf{z}) & = \sum_{j=1}^n \log \left (\sum_{t=1}^s w_t (\bX_j) e^{z_t} \right)\\
& = \sum_{j=1}^n \log \left ( f(\bX_j) \sum_{k=1}^d \I \left \{ X_{jk} \geq \gamma \right \} e^{z_1} + f_2(\bX_j)e^{z_2} \right . \\
& \qquad \qquad \qquad + f(\bX_j)\I \left \{ S(\bX_j) \geq \gamma, M(\bX_j) < \gamma \right \} e^{z_3}   \\
& \qquad \qquad \qquad + f(\bX_j) \I \{S(\bX_j) \geq \gamma \} e^{z_4} \Bigg  )\\
& = \sum_{j=1}^n \log \left (\sum_{k=1}^d \I \left \{ X_{jk} \geq \gamma \right \} e^{z_1}+ \frac{f_2(\bX_j)}{f(\bX_j)}e^{z_2} \right . \\
& \qquad \qquad \qquad + \I \left \{ S(\bX_j) \geq \gamma, M(\bX_j) < \gamma \right \} e^{z_3}  \\
& \qquad \qquad \qquad +  \I \{S(\bX_j) \geq \gamma \} e^{z_4}   \Bigg ) + \sum_{j=1}^n \log f(\bX_j). 
\end{align*}
Hence, solving this will allow us to obtain an optimal solution $\widehat{\mathbf{z}}$, and $\widehat{\Bell}$ up to a multiplicative constant. Using the analytical knowledge of known $\ell_1$ or $\ell_2$, we are able to recover the remaining $\widehat{\Bell}$. Thus, we are able to obtain the estimator of the (rare-event) probability of interest $\widehat{\ell}_s$.
For implementation purposes, we can use the following lemma to speed up the computation time in evaluating the product of marginals from the zero-variance density~\eqref{eq:min_var_IS}.

\begin{Lemma} \label{lemma:f2_simplify}
The $i$th marginal of the zero-variance density can be written as 
\begin{equation} 
\widehat{\pi}_i (x_i) = \frac{f(x_i)}{n_s} \sum_{j=1}^{n_s} e^{C_{ji}^{\alpha}} \I \{x_i \geq C_{ji} \},
\end{equation}
where $$C_{ji}= \max\left\{0,\gamma-\sum_{k\neq i} X_{jk}\right\}, $$
are calculated constants from samples of $f_s$.
It follows that
\begin{equation} \label{f2_simplify}
\frac{f_2(\bx)}{f(\bx)}= \prod_{i=1}^d \frac{1}{n_s}\sum_{j=1}^{n_s}  e^{C_{ji}^{\alpha}} \I \{x_i \geq C_{ji}\}, \ \bx=(x_1,\dots,x_d). 
\end{equation}
\end{Lemma}

\begin{proof}
The proof of lemma~\reff{ELS_Scheme1}{lemma:f2_simplify} has been delegated to appendix~\ref{appendix:proof_f2_simplify}.
\end{proof}
The computational cost measured by CPU time will depend on its implementation in software. An efficient implementation to evaluate~\eqref{f2_simplify} can be achieved by sorting the vector $[C_{ji}], j=1,\dots,n_s$, and pre-computing the cumulative sum $\sum_{j=1}^{k} e^{C_{ji}^{\alpha}}$, for each $k=1,\dots,n_s$. This is followed by simply looking up which index boundaries the samples fall into. Using vectorization and suitable index `look-up' functions may greatly speed up the computational process. For example, the {\tt histc} function in {\tt MATLAB} outputs the bin count which is an efficient way to look up the values in~\eqref{f2_simplify}. See page~\pageref{appendix:matlab_ELM_scheme1} of the appendix for this implementation in {\tt MATLAB}.

We can improve the accuracy of each marginal density $\widehat{\pi}_i(x_i)$, by exploiting symmetry and iid properties of our samples from $f_s$. This is achieved by increasing the sample size used from $n_s$ to  $n_s \times d$. By utilizing all elements of each random sample, a better estimator is
$$\widehat{\pi} (x_i) = \frac{f(x_i)}{n_s d} \sum_{i=1}^d\sum_{j=1}^{n_s} e^{C_{ji}^{\alpha}} \I \{x_i \geq C_{ji} \}. $$
Further, it also possible to reduce the computational costs by taking a sub-sample of existing $\bX_1,\dots,\bX_s$ to cut down the number of summation terms in~\eqref{f2_simplify} but this will reduce the estimator's accuracy. There exists a trade-off between variance reduction and computational speed.

\section{Numerical Results with Discussion A} \label{numerical_results1}
For the problem stated in section~\ref{problem_statement},  comprehensive simulations were run to  compare the performance of the ELM estimator against the MCIS~\eqref{eq:MCIS_estimator} and the AK conditional estimator~\eqref{eq:AK_estimator}. 
Firstly, the shape parameter $\alpha$ is varied from $0.1$ to $0.9$ in order to illustrate changes displayed by the estimators as the distribution \textsf{Weib($\alpha,1$)} becomes less heavy-tailed when $\alpha$ approaches $1$.
Similarly, the rarity parameter $\gamma$ is increased such that the probability of the rare-event becomes smaller. The aim is to capture any numerical display of second order efficiency properties (see appendix~\ref{appendix:efficiency}) by the estimators. 

The numerical package {\tt MATLAB} is used to implement the given ELM schemes. The function {\tt fmincon} from the optimization toolbox is used to solve the constrained nonlinear convex optimization program. Simulations were ran on a quad core, 2.83GHz computer. 
See appendix~\ref{appendix:matlab_ELM_scheme1} for the main program and functions used in {\tt MATLAB}.
The relative error~\eqref{RE_formula} and relative time variance product~\eqref{RTVP_formula} for each of the estimators are given. 
These are key quantities used to compare efficiency between the estimators.
The estimate $\widehat{\ell}$ (to three significant figures) and the average CPU time (in seconds) are also given.

We begin by examining the results from the ELM scheme with 4 densities. 
Further, to illustrate the trade off between variance reduction and computational speed, we also give results for an ELM scheme with 3 densities ($f_3$ discarded).
For a fair comparison, the sample sizes used to compute the MCIS and AK conditional estimators are adjusted accordingly.

\subsection{ELM Scheme with 4 densities}

\paragraph{Parameters:} We consider the case when $d=10$. Equal sample sizes of $n_t=10^4$ from each density $f_t$, $t=1,\dots,4$ are used. Hence, the pooled sample size is $n=4\times 10^{4}$. A random sub-sample of size $0.5\times n_s$ is taken from $\bX_1,\dots,\bX_{n_s}$ to estimate the product of marginals $f_2$. We have used $K=100$ iterations of the ELM algorithm to estimate its empirical standard error.
Both the MCIS and AK conditional estimators used an equivalent sample size of $n=4\times 10^4$.
The MCIS estimator used sample size of $m=0.5\times n$ to estimate the product of marginals importance sampling density.

\paragraph{Discussion of Results:} 
The numerical results of the ELM scheme are given in Table~\ref{tab:sampling_scheme1} below. 
We observe that the ELM is superior to the MCIS estimator with overall smaller relative error and RTVP.  
The ELM estimator performed better than the AK conditional estimator with smaller relative error and RTVP when $\alpha=0.6,0.8,0.9$. 
Similarly, the MCIS estimator had lower RTVP than the AK conditional estimator when $\alpha=0.8,0.9$. 
However, the AK conditional estimator outperforms both the ELM and MCIS estimators when $\alpha=0.1,0.2$.  
These results are expected as it is known that the AK conditional estimator performs very well when $0<\alpha<0.369$.

Further, for $\alpha=0.1,0.2$, the AK conditional estimator is observed to enjoy the property vanishing relative error properties when $\gamma \rightarrow \infty$ (so that $\ell \downarrow 0$). 
Similarly, we observe that the ELM and MCIS estimators display numerical bounded relative error properties where $\text{RE}(\widehat{\ell}(\gamma)) < K$ for some $K$ independent of $\gamma$.  
The performance of the ELM estimator seems to be consistent for the whole range of $\alpha<1$ with relative error of order $10^{-3}$.
This is also true for the MCIS estimator with relative error of order $10^{-2}$.
All three estimators have smaller relative error and RTVP when $\alpha$ is small. This suggests these estimators are more efficient when the distribution is more heavy-tailed.

On average, the computational time required for the ELM method is 6.6 times higher than the MCIS procedure and 2.6 times higher than computing the AK conditional estimator.
After close examination of the ELM procedure, we found the Gibbs sampling procedure for $f_3$ to be the culprit for the bottleneck in computational speed. 
It is worthwhile to investigate whether discarding $f_3$ will improve the RTVP for the ELM procedure. 
Hence, we perform another simulation run of the same ELM scheme but with $f_3$ removed.

\subsection{ELM Scheme with 3 densities}

\paragraph{Parameters:} 

The density that captures the residual component of the rare-event ($f_3$) has been omitted.
Hence, the overall sample size has been reduced to $n=3\times 10^4$ for each of the estimation methods. Other algorithmic parameters have remained the same.

\paragraph{Discussion of Results:} 
Table~\ref{tab:sampling_scheme2} shows the results from another simulation study using 3 densities instead of 4 in the ELM scheme ($f_3$ discarded).
We observe there is an overall increase in the relative error for the ELM estimator. 
The is due to the removal of $f_3$ in the ELM procedure.
During the initial runs of searching suitable ELM densities, we have found that the inclusion of $f_3$ contributes to some degree of variance reduction.

However, by removing such density, we obtain significant gains in computational speed (average CPU time of $0.35$s compared with $1.92$s). The bottleneck from the original scheme of 4 densities seem to result from sampling from such density.  
On average, the ELM scheme with 3 densities is approximately 5.5 times faster than the ELM scheme with 4 densities.
Further, by comparing average CPU time, the ELM procedure now runs faster than computing the AK conditional estimator (0.54s) but slower than calculating the MCIS estimator (0.24s).

This ELM estimator has obtained significant gains in efficiency.
Overall, by comparing similar levels of $\alpha$ and $\gamma$, the RTVP for the ELM with 3 densities is smaller than the ELM procedure with 4 densities.
This suggests the variance reduction achieved by adding in $f_3$ is not high enough to warrant the increase in computational time.  
Due to a decrease in sample size, the computational time is also slightly reduced for obtaining the MCIS and AK estimators. 
Note we still observe similar patterns in the relative performance of the three estimators.

\begin{table}[H]
	\centering
  \caption{Comparison of the ELM method (4 densities) with the MCIS and AK conditional estimator for different $\alpha$ and $\gamma$. }
	\footnotesize
		\begin{tabular}{|c|c|rrr|rrr|}
    \hline
		  \multicolumn{2}{|c|}{$\alpha=0.9$}    & \multicolumn{3}{c|}{Relative Error} &       & RTVP  &  \\
		\hline
    $\gamma$ & $\widehat{\ell}$ &  ELM   & MCIS  & AK    & ELM   & MCIS  & AK \\
    \hline
    30    & $1.33/10^4$   & $1.66/10^3$ & $1.29/10^2$ & $5.73/10^2$ & $6.45/10^6$ & $4.61/10^5$ & $2.33/10^3$ \\
		40    & $6.25/10^7$   & $2.00/10^3$ & $1.68/10^2$ & $3.24/10^1$ & $8.00/10^6$ & $9.15/10^5$ & $7.91/10^2$ \\
    50    & $2.25/10^9$   & $2.27/10^3$ & $1.86/10^2$ & $2.99/10^1$ & $1.12/10^5$ & $1.16/10^4$ & $6.76/10^2$ \\
    60    & $7.01/10^{12}$  & $2.70/10^3$ & $2.26/10^2$ & $2.98/10^1$ & $1.47/10^5$ & $1.73/10^4$ & $6.64/10^2$ \\
    90    & $1.54/10^{19}$  & $3.33/10^3$ & $2.86/10^2$ & $3.92/10^1$ & $2.17/10^5$ & $2.68/10^4$ & $1.16/10^1$ \\
    \hline
    \hline
    \multicolumn{2}{|c|}{$\alpha=0.8$}   &  \multicolumn{3}{c|}{Relative Error} &       & RTVP  &  \\
\hline
    $\gamma$ & $\widehat{\ell}$     & ELM   & MCIS  & AK    & ELM   & MCIS  & AK \\
    \hline
    30    & $1.59/10^3$   & $1.61/10^3$ & $1.17/10^2$ & $2.29/10^2$ & $4.91/10^6$ & $4.38/10^5$ & $3.96/10^4$ \\
    50    & $1.02/10^6$   & $2.01/10^3$ & $1.72/10^2$ & $1.03/10^1$ & $7.95/10^6$ & $9.17/10^5$ & $7.93/10^3$ \\
    100   & $1.44/10^{14}$  & $3.31/10^3$ & $2.65/10^2$ & $2.16/10^1$ & $2.15/10^5$ & $2.19/10^4$ & $3.47/10^2$ \\
    150   & $1.33/10^{21}$  & $3.53/10^3$ & $4.10/10^2$ & $1.12/10^1$ & $2.54/10^5$ & $5.08/10^4$ & $9.35/10^3$ \\
    200   & $5.17/10^{28}$   & $3.04/10^3$ & $2.37/10^2$ & $5.99/10^2$ & $1.89/10^5$ & $1.84/10^4$ & $2.68/10^3$ \\
    \hline
      \hline
    \multicolumn{2}{|c|}{$\alpha=0.6$}     & \multicolumn{3}{c|}{Relative Error} &       & RTVP  &  \\
\hline
  $\gamma$ & $\widehat{\ell}$ & ELM   & MCIS  & AK    & ELM   & MCIS  & AK \\
    \hline
    100   & $9.47/10^6$& $2.02/10^3 $& $1.49/10^2 $& $2.11/10^2 $& $7.65/10^6 $&$ 6.22/10^5 $& $3.20/10^4 $\\
    150   & $7.86/10^8$ & $2.35/10^3 $&$ 1.56/10^2 $& $3.63/10^2 $& $9.26/10^6 $& $6.12/10^5 $& $9.30/10^4 $\\
    200   & $1.35/10^9$& $1.74/10^3 $& $1.64/10^2 $& $1.16/10^2 $& $5.12/10^6 $& $8.41/10^5 $& $1.03/10^4 $\\
    500   & $1.84/10^{17}$& $1.03/10^3$ & $1.21/10^2 $& $3.59/10^3$ & $2.29/10^6 $&$ 3.72/10^5 $&$ 9.29/10^6 $\\
    1000  & $7.01/10^{27}$ & $1.03/10^3$ & $1.17/10^2 $& $2.09/10^3$ & $1.95/10^6 $& $3.47/10^5 $& $3.16/10^6$ \\
    \hline
    \hline
    \multicolumn{2}{|c|}{$\alpha=0.2$}    & \multicolumn{3}{c|}{Relative Error} &       & RTVP  &  \\
\hline
   $\gamma$ & $\widehat{\ell}$  & ELM   & MCIS  & AK    & ELM   & MCIS  & AK \\
    \hline
    $10^4$  & $1.97/10^2$  & $4.42/10^4 $&$ 7.81/10^3 $& $1.02/10^3 $&$ 3.29/10^7 $&$ 1.56/10^5 $&$ 7.63/10^7 $\\
    $10^5$  & $4.64/10^4$  & $3.41/10^4 $& $6.97/10^3 $& $5.34/10^4 $& $2.37/10^7 $& $1.17/10^5 $&$ 2.05/10^7 $\\
    $10^6$  & $1.31/10^6$   & $3.83/10^4 $& $6.64/10^3 $& $8.38/10^5 $& $2.59/10^7 $& $1.18/10^5 $&$ 5.08/10^9 $\\
    $10^7$  & $1.23/10^{10}$  &$ 4.01/10^4 $&$ 6.33/10^3$ &$ 2.08/10^5$ &$ 2.83/10^7$ &$ 1.05/10^5 $&$ 3.19/10^{10} $\\
    $10^8$  & $5.13/10^{17}$  &$ 3.51/10^4$ & $6.35/10^3$ & $1.97/10^6$ & $2.43/10^7$ & $9.95/10^6 $& $2.87/10^{12}$\\
    \hline
		\hline
   	\multicolumn{2}{|c|}{$\alpha=0.1$}   & \multicolumn{3}{c|}{Relative Error} &       & RTVP  & \\
\hline
    $\gamma$ & $\widehat{\ell}$   & ELM   & MCIS  & AK    & ELM   & MCIS  & AK\\
    \hline
    $10^{10}$ & $4.54/10^4$   & $3.57/10^4 $&$ 6.26/10^3 $&$ 1.27/10^4 $&$ 2.17/10^7 $&$ 1.06/10^5 $&$ 1.20/10^8 $\\
    $10^{11}$ & $3.41/10^5$ & $3.37/10^4 $& $6.30/10^3 $&$ 5.72/10^5 $&$ 2.12/10^7 $& $1.12/10^5 $&$ 2.37/10^9 $\\
    $10^{12}$ & $1.31/10^6$   & $3.41/10^4 $& $6.29/10^3 $& $6.52/10^6 $& $2.45/10^7 $& $1.02/10^5 $&$ 3.02/10^{11} $\\
    $10^{13}$ & $2.16/10^8$  & $3.36/10^4 $& $6.35/10^3 $& $1.54/10^6 $& $1.94/10^7 $& $1.01/10^5 $&$ 1.69/10^{12} $\\
    $10^{15}$ & $1.85/10^{13}$  &$ 3.10/10^4$ &$ 6.33/10^3$ & $7.35/10^8$ &$ 1.81/10^7$ &$ 1.10/10^5$ & $3.87/10^{15}$\\
    \hline
		    \multicolumn{1}{r}{} & \multicolumn{1}{r}{} & \multicolumn{1}{r}{} & \multicolumn{1}{r}{} & \multicolumn{1}{r}{} &       &       & \multicolumn{1}{r}{} \\
\cline{3-5}    \multicolumn{1}{r}{} &       & \multicolumn{3}{c|}{Average CPU Time (s)} &       &       & \multicolumn{1}{r}{} \\
\cline{3-5}    \multicolumn{1}{r}{} &       & ELM   & MCIS  & AK    &       &       & \multicolumn{1}{r}{} \\
    \multicolumn{1}{r}{} &       & 1.92  & 0.29  & 0.73  &       &       & \multicolumn{1}{r}{} \\
\cline{3-5}    
		\end{tabular}%
  \label{tab:sampling_scheme1}%
\end{table}%


\begin{table}[H]
  \centering
  \caption{Comparison of the ELM method (3 densities) with the MCIS and AK conditional estimator for different $\alpha$ and $\gamma$}
		\footnotesize
    \begin{tabular}{|r|r|r|r|r|rrr|}
    \hline
    \multicolumn{2}{|c|}{$\alpha=0.9$} & \multicolumn{3}{c|}{Relative Error} &       & RTVP  &  \\
    \hline
    $\gamma$ & $\widehat{\ell}$ & \multicolumn{1}{r}{ELM} & \multicolumn{1}{r}{MCIS} & AK    & ELM   & MCIS  & AK \\
    \hline
    $30 $   & $1.33/10^4$ & \multicolumn{1}{r}{$1.84/10^3$} & \multicolumn{1}{r}{$1.51/10^2$} & $7.69/10^2$ & $3.83/10^6$ & $5.78/10^5$ & $3.16/10^3$ \\
    $40 $   & $6.25/10^7$ & \multicolumn{1}{r}{$2.13/10^3$} & \multicolumn{1}{r}{$1.82/10^2$} & $2.11/10^1$ & $1.31/10^6$ & $8.30/10^5$ & $2.37/10^2$ \\
    $50 $   & $2.25/10^9$ & \multicolumn{1}{r}{$2.52/10^3$} & \multicolumn{1}{r}{$2.18/10^2$} & $9.46/10^1$ & $1.94/10^6$ & $1.20/10^4$ & $4.77/10^1$ \\
    $60 $   & $7.01/10^{12}$ & \multicolumn{1}{r}{$2.75/10^3$} & \multicolumn{1}{r}{$2.50/10^2$} & $2.74/10^1$ & $3.18/10^6$ & $1.59/10^4$ & $4.01/10^2$ \\
    $90 $   & $1.54/10^{19}$ & \multicolumn{1}{r}{$4.49/10^3$} & \multicolumn{1}{r}{$3.36/10^2$} & $8.91/10^1$ & $6.82/10^6$ & $2.85/10^4$ & $4.23/10^1$ \\
    \hline
    \hline
    \multicolumn{2}{|c|}{$\alpha=0.8$} & \multicolumn{3}{c|}{Relative Error} &       & RTVP  &  \\
    \hline
    $\gamma$ & $\widehat{\ell}$ & \multicolumn{1}{r}{ELM} & \multicolumn{1}{r}{MCIS} & AK    & ELM   & MCIS  & AK \\
    \hline
    $30 $   & $1.59/10^3$ & \multicolumn{1}{r}{$1.66/10^3$} & \multicolumn{1}{r}{$1.37/10^2$} & $2.69/10^2$ & $8.20/10^7$ & $4.58/10^5$ & $3.88/10^4$ \\
    $50 $   & $1.02/10^6$ & \multicolumn{1}{r}{$2.50/10^3$} & \multicolumn{1}{r}{$1.98/10^2$} & $1.18/10^1$ & $2.05/10^6$ & $9.66/10^5$ & $7.53/10^3$ \\
    $100$   & $1.44/10^{14}$ & \multicolumn{1}{r}{$4.11/10^3$} & \multicolumn{1}{r}{$2.99/10^2$} & $3.86/10^1$ & $6.03/10^6$ & $2.19/10^4$ & $7.97/10^2$ \\
    $150$   & $1.33/10^{21}$ & \multicolumn{1}{r}{$4.03/10^3$} & \multicolumn{1}{r}{$4.84/10^2$} & $1.60/10^1$ & $6.31/10^6$ & $5.68/10^4$ & $1.37/10^2$ \\
    $200$   & $5.17/10^{28}$ & \multicolumn{1}{r}{$3.17/10^3$} & \multicolumn{1}{r}{$3.42/10^2$} & $5.40/10^1$ & $4.52/10^6$ & $2.82/10^4$ & $1.55/10^1$ \\
    \hline
    \hline
    \multicolumn{2}{|c|}{$\alpha=0.6$} & \multicolumn{3}{c|}{Relative Error} &       & RTVP  &  \\
    \hline
    $\gamma$ & $\widehat{\ell}$ & \multicolumn{1}{r}{ELM} & \multicolumn{1}{r}{MCIS} & AK    & ELM   & MCIS  & AK \\
    \hline
    $100$   & $9.47/10^6$ & \multicolumn{1}{r}{$2.13/10^3$} & \multicolumn{1}{r}{$1.83/10^2$} & $2.67/10^2$ & $1.31/10^6$ & $7.76/10^5$ & $3.82/10^4$ \\
    $150$   & $7.86/10^8$ & \multicolumn{1}{r}{$2.06/10^3$} & \multicolumn{1}{r}{$1.81/10^2$} & $1.95/10^2$ & $1.25/10^6$ & $7.75/10^5$ & $2.03/10^4$ \\
    $200$   & $1.35/10^9$ & \multicolumn{1}{r}{$1.99/10^3$} & \multicolumn{1}{r}{$2.22/10^2$} & $4.33/10^2$ & $1.13/10^6$ & $1.26/10^4$ & $1.01/10^3$ \\
    $500$   & $1.84/10^{17}$ & \multicolumn{1}{r}{$1.18/10^3$} & \multicolumn{1}{r}{$1.63/10^2$} & $3.90/10^3$ & $4.72/10^7$ & $6.03/10^5$ & $8.12/10^6$ \\
    $1000$  & $7.01/10^{27}$ & \multicolumn{1}{r}{$9.17/10^4$} & \multicolumn{1}{r}{$1.41/10^2$} & $2.65/10^3$ & $2.79/10^7$ & $4.78/10^5$ & $3.89/10^6$ \\
    \hline
    \hline
    \multicolumn{2}{|c|}{$\alpha=0.2$} & \multicolumn{3}{c|}{Relative Error} &       & RTVP  &  \\
    \hline
    $\gamma$ & $\widehat{\ell}$ & \multicolumn{1}{r}{ELM} & \multicolumn{1}{r}{MCIS} & AK    & ELM   & MCIS  & AK \\
    \hline
    $10^4$  & $1.97/10^2$ & \multicolumn{1}{r}{$4.55/10^4$} & \multicolumn{1}{r}{$8.83/10^3$} & $1.17/10^3$ & $5.98/10^8$ & $1.78/10^5$ & $7.71/10^7$ \\
    $10^5$  & $4.64/10^4$ & \multicolumn{1}{r}{$3.98/10^4$} & \multicolumn{1}{r}{$7.85/10^3$} & $6.45/10^4$ & $5.32/10^8$ & $1.40/10^5$ & $2.34/10^7$ \\
    $10^6$  & $1.31/10^6$ & \multicolumn{1}{r}{$3.40/10^4$} & \multicolumn{1}{r}{$7.48/10^3$} & $2.14/10^4$ & $4.07/10^8$ & $1.31/10^5$ & $2.56/10^8$ \\
    $10^7$  & $1.23/10^{10}$ & \multicolumn{1}{r}{$3.46/10^4$} & \multicolumn{1}{r}{$7.35/10^3$} & $1.65/10^5$ & $3.94/10^8$ & $1.26/10^5$ & $1.53/10^{10}$ \\
    $10^8$  & $5.13/10^{17}$ & \multicolumn{1}{r}{$3.46/10^4$} & \multicolumn{1}{r}{$7.33/10^3$} & $2.54/10^6$ & $4.53/10^8$ & $1.20/10^5$ & $3.54/10^{12}$ \\
    \hline
    \hline
    \multicolumn{2}{|c|}{$\alpha=0.1$} & \multicolumn{3}{c|}{Relative Error} &       & RTVP  &  \\
    \hline
    $\gamma$ & $\widehat{\ell}$ & \multicolumn{1}{r}{ELM} & \multicolumn{1}{r}{MCIS} & AK    & ELM   & MCIS  & AK \\
    \hline
    $10^{10}$ & $4.54/10^4$ & \multicolumn{1}{r}{$3.73/10^4$} & \multicolumn{1}{r}{$7.29/10^3$} & $1.23/10^4$ & $6.67/10^8$ & $1.19/10^5$ & $8.29/10^9$ \\
    $10^{11}$ & $3.41/10^5$ & \multicolumn{1}{r}{$3.21/10^4$} & \multicolumn{1}{r}{$7.29/10^3$} & $6.88/10^5$ & $4.28/10^8$ & $1.19/10^5$ & $2.62/10^9$ \\
    $10^{12}$ & $1.31/10^6$ & \multicolumn{1}{r}{$3.65/10^4$} & \multicolumn{1}{r}{$7.25/10^3$} & $8.07/10^6$ & $4.42/10^8$ & $1.17/10^5$ & $3.59/10^{11}$ \\
    $10^{13}$ & $2.16/10^8$ & \multicolumn{1}{r}{$3.61/10^4$} & \multicolumn{1}{r}{$7.25/10^3$} & $5.24/10^7$ & $3.70/10^8$ & $1.14/10^5$ & $1.51/10^{13}$ \\
    $10^{15}$ & $1.85/10^{13}$ & \multicolumn{1}{r}{$3.50/10^4$} & \multicolumn{1}{r}{$7.22/10^3$} & $2.01/10^8$ & $4.57/10^8$ & $1.15/10^5$ & $2.22/10^{16}$ \\
    \hline
    \multicolumn{1}{r}{} & \multicolumn{1}{r}{} & \multicolumn{1}{r}{} & \multicolumn{1}{r}{} & \multicolumn{1}{r}{} &       &       & \multicolumn{1}{r}{} \\
\cline{3-5}    \multicolumn{1}{r}{} &       & \multicolumn{3}{c|}{Average CPU Time (s)} &       &       & \multicolumn{1}{r}{} \\
\cline{3-5}    \multicolumn{1}{r}{} &       & ELM   & MCIS  & AK    &       &       & \multicolumn{1}{r}{} \\
    \multicolumn{1}{r}{} &       & 0.35  & 0.24  & 0.54  &       &       & \multicolumn{1}{r}{} \\
\cline{3-5}    \end{tabular}%
  \label{tab:sampling_scheme2}%
\end{table}%

\newpage

\section{ELM Scheme B: 2 densities with inequality constraint} \label{ELM_scheme_B}
In section~\ref{constrained_MEL}, we have mentioned that by maximizing the empirical likelihood
subject to both equality and inequality constraints, the efficiency of the estimator  $\widehat\bz$ can be improved. 
In this section, we present a new variational optimization method for obtaining a lower bound for rare-event probabilities of interest.
The method not only provides a lower bound on $\ell$, which can be incorporated into the likelihood optimization, but as a byproduct it provides an excellent reference density to be used in the ELM sequence of densities.  We now explain how we obtain these lower bounds before giving the details of the ELM algorithm that exploits the lower bound.  The main finding is summarized in Theorem~\reff{ELM_scheme_B}{lower_bound_lemma} below.

\subsection{Variational Lower Bound on $\ell$} \label{lower_bound}

Here, we introduce a novel method in obtaining a lower bound for our (rare-event) probability $\ell$.
This serves as the key ingredient in implementing ELM scheme B in section~\ref{ELS_Scheme2} below. 
Firstly, we need the following observation to rewrite our $\textsf{Weib}(\alpha,1)$ random variables in terms of $\textsf{Exp}(1)$ random variables.
The (rare-event) probability $$\ell=\Pm(X_1+\cdots+X_d\geq\gamma),$$ with $X_i\simiid\textsf{Weib}(\alpha,1)$ and $0 < \alpha < 1$, is equivalent to 
\[
\ell=\Pm(Y_1^{1/\alpha}+\cdots+Y_d^{1/\alpha}\geq\gamma),\quad Y_i\simiid\textsf{Exp}(1)\;.
\]
Now, we use the following theorem to establish a lower bound $\ell_{L}$ for the (rare-event) probability of interest $\ell$. That is, $\ell_{L} \leq \ell.$
This will be used to construct a reference density in section~\ref{ELS_Scheme2} below.

\begin{Theorem} \label{lower_bound_lemma}
For any positive value of the variational parameters $\lambda_1,\ldots,\lambda_d$, we have a lower bound to $\ell$  given by
\begin{equation}
\ell_{L} =  \Pm(S_L(\bY;\boldsymbol \lambda)\geq\gamma) = Q\left(\boldsymbol\beta;\alpha\gamma+(1-\alpha)\sum_i\lambda_i^{1/\alpha}\right),
\end{equation}
where
\begin{equation}
S_L(\by;\boldsymbol \lambda) \stackrel{def}{=} \frac{1}{\alpha}\sum_i \lambda_i^{1/\alpha-1} y_i-\frac{1-\alpha}{\alpha}\sum_i \lambda_i^{1/\alpha}
\end{equation} 
and the function $Q(\boldsymbol\beta;\gamma)=(1,0,\ldots,0) \;\e^{A\,\gamma}\; \mathbf{1}$ is defined via the matrix exponential with
\[
A=\left(\begin{array}{ccccc}
-\beta_1 & \beta_1 & 0 & \cdots &0 \\
0 & -\beta_2 & \beta_2 & \cdots &0 \\
\vdots & \vdots & \ddots & \ddots &\vdots \\
0 & \cdots & 0 & -\beta_{d-1} &\beta_{d-1}  \\
0 & \cdots & 0 & 0 &-\beta_{d}  \\
 \end{array}\right).
\]
and 
$$\beta_j=\frac{\lambda_j^{1/\alpha-1}+\cdots+\lambda_d^{1/\alpha-1}}{d-j+1}, \qquad j=1,\dots,d.$$
\end{Theorem}

\begin{proof}
The proof of this theorem is given in appendix~\ref{lower_bound_lemma_proof}.
\end{proof}

Note the function $Q(\boldsymbol\beta;\gamma)$ is the tail probability of the \emph{generalized Erlang distribution}~\cite[\S 4.2.13]{kroese2011handbook}. 
It is the distribution of a linear combination of $d$ independent exponential random variables with equal or unequal rates.
This quantity can be can be calculated analytically. 
Ideally, we want to to obtain a lower bound probability $\ell_L$ close  to the true (rare-event) probability.
This can be achieved by solving the following maximization program
\begin{equation} \label{lower_bound_maximization}
\boldsymbol \lambda^{*} =  \argmax_{\boldsymbol\lambda > \boldsymbol 0} \ \Pm(S_L(\bY;\boldsymbol \lambda) \geq \gamma).
\end{equation}
Here $\boldsymbol \lambda^{*}$ is the optimal value of the variational parameters  that maximizes the lower bound probability.
Hence, we can establish a ELM scheme which uses a reference density whose normalizing constant represents the lower bound of the (rare-event) probability of interest. 

\subsection{Formulation of Scheme B}
\label{ELS_Scheme2}
Firstly, we find the optimal solution  $\boldsymbol \lambda^{*}$  to the maximization program given by~\eqref{lower_bound_maximization}.
This optimization problem may be difficult to solve as it is nonlinear and not convex. 
However, any heuristic approach such as the cross-entropy method for optimization~\cite[\S 13.2]{kroese2011handbook} would work. 
Maximizing this probability will give a lower bound to the probability of interest $\ell$ which can be incorporated in the ELM scheme.

We apply the procedure as described in section~\ref{sec:ELS_formulation} and~\ref{constrained_MEL} with $s=2$ densities. 
Suppose that $f(\bx)$ is the joint density of iid \textsf{Exp}$(1)$ distribution. Consider the following densities
\begin{enumerate}
\item Lower bound reference density
\[ f_1(\bx) = \frac{f(\bx)\ \I \left \{ x_{[1]} \lambda_1^{*(1/\alpha-1)}+\cdots+x_{[d]} \lambda_d^{*(1/\alpha-1)} \geq \gamma^{*} \right\}}{\ell_1}:=\frac{w_1(\bx) }{\ell_1}, \]
where
\[\gamma^{*} = \alpha\gamma+(1-\alpha)\sum_i\lambda_i^{*1/\alpha}, \]
and $x_{[1]},x_{[2]},\dots,x_{[d]}$ represents the order  statistics. Here, $\ell_1$ is known analytically and is given by
$$\ell_1 = Q\left(\boldsymbol\beta^{*};\gamma^{*}\right),$$
with
$$\beta^*_j=\frac{\lambda_j^{*1/\alpha-1}+\cdots+\lambda_d^{*1/\alpha-1}}{d-j+1}, \qquad j=1,\dots,d.$$
An important point is that while the ELM can exploit this lower bound reference density, the reference density by itself \textbf{cannot} be used as an importance sampling density.

\item Zero-variance density
\[ f_2(\bx) = \frac{f(\bx)\ \I \left \{ x_{[1]}^{1/\alpha}+\cdots+x_{[d]}^{1/\alpha} \geq \gamma \right\}}{\ell_2}:=\frac{w_2(\bx) }{\ell_2}, \]
where $\ell_2$ is the unknown (rare-event) probability of interest.
\end{enumerate}
Again we emphasize an important advantage of the ELM method:  since  the reference density $f_1$ exhibits light tailed behavior (see appendix~\ref{appendix:tail_properties}) and its support is a subset of the support of $f_s$, it cannot be used as an importance sampling density to estimate $f_s$. Nevertheless, it makes for an excellent ELM density.


\subsection{Sampling}
For this scheme there is no need to generate from $f_1$. Note that $$\{S_L(\bY;\boldsymbol \lambda)\geq\gamma\}\subseteq\{S(\bY)\geq\gamma\}, $$ and thus samples attained from $f_1$ and $f_2$ always satisfies the condition imposed by the zero variance density
$$\I \left (x_{[1]}^{1/\alpha}+\cdots+x_{[d]}^{1/\alpha} \geq \gamma \right). $$
However, we will need to sample from $f_1$ if we change this ELM, say, by adding another reference density. To proceed, we will need to use the Gibbs Sampler to generate samples from 
$$g(\bz) = \frac{f(\bx) \I \left \{ \sum_{j=1}^d z_j \beta^*_j \geq \gamma^* \right \}}{\ell_1}.$$
In the Gibbs Sampler, generating from the conditional distribution
$$ g(z_i | \bz_{-i}) \propto f(z_i), \qquad \max \left (0,\frac{1}{\beta_i^*}\left[\gamma^* - \sum_{j \neq i} z_j \beta^*_j\right] \right) \leq z_i < \infty, $$
can be achieved by setting the $i$th component as
$$ z_i := \left [ \max \left (0,\frac{1}{\beta_i^*}\left[\gamma^* - \sum_{j \neq i} z_j \beta^*_j\right] \right) - \log U \right ], \qquad U \sim \textsf{Uniform}(0,1).$$
Finally, realizations from $f_1$ can be attained by setting the $i$th component as
$$x_{[i]} := \sum_{j=1}^i \frac{z_j}{d-j+1}.$$
Sampling from $f_2$ is straightforward by using the Gibbs Sampler and then sorting the components of generated $\bx_j$'s in increasing order.

\subsection{Empirical Likelihood Optimization}
Now, parametrize
$$\bz = (z_1,z_2) = (-\log(\ell_1/\tilde{\lambda}_1),-\log(\ell_2/\tilde{\lambda}_2)), $$
where $\tilde{\lambda}_1$ and $\tilde{\lambda}_2$ are the proportion of samples attained from $f_1$ and $f_2$ respectively.
The objective function is
\begin{align*}
\widehat{D}(\mathbf{z}) & = \sum_{j=1}^n \log \left (\sum_{t=1}^2 w_t (\bX_j) e^{z_t} \right).
\end{align*}
Note, in this case, $\widehat{D}(\mathbf{z})$ can be simplified, because it depends on the   sufficient statistic  $\widehat p$, where $\widehat p$ denotes the number of samples $\bX_j$'s that satisfy  
$$X_{[1]} \lambda_1^{*(1/\alpha-1)}+\cdots+X_{[d]} \lambda_d^{*(1/\alpha-1)} \geq \gamma^{*}.$$
Then, we have
\begin{align*}
\widehat{D}(\mathbf{z}) & = \sum_{j=1}^n \log \left (\sum_{t=1}^2 w_t (\bX_j) e^{z_t} \right) \\
& = \underbrace{\widehat p \log(\e^{z_1}+\e^{z_2}) + (n-\widehat p) z_2}_{\stackrel{\mathrm{def}}{=}  \widehat{Q}(\mathbf{z})} + \underbrace{\sum_{j=1}^n \log f(\bX_j)}_{\mathrm{constant}} 
\end{align*}
We can then minimize  $\widehat{Q}(\mathbf{z})$ instead of $\widehat{D}(\mathbf{z})$. 
Hence, we solve the following convex optimization program to obtain $\widehat{\bz}$ and thus $\widehat{\ell}$ up to a multiplicative constant. 
\begin{align*}
& \widehat{\bz} = \min_{\bz} \widehat{Q}(\bz)\\
& \text{subject to:}\\
& \tilde{\lambda}_1 z_1 + \tilde{\lambda}_2 z_2 = 0, \\
& -z_1 + z_2 \leq -\log(\tilde{\lambda}_1) + \log(\tilde{\lambda}_2).
\end{align*}
Using knowledge of $\ell_1$, we can recover an estimate $\widehat{\ell}_2$ of the (rare-event) probability. 

We require the equality constraint to obtain an unique solution from the parameter size.
The inequality constraint allows us to use the analytical knowledge that $\ell_1 \leq \ell_2.$

\section{Numerical Results with Discussion B} \label{numerical_results2}
Similar to ELM scheme A, we perform comprehensive simulations using ELM scheme B for the problem stated in section~\ref{problem_statement}.
We vary the shape parameter $\alpha$ from $0.1$ to $0.9$ in order to capture any changes displayed by the estimators as $\alpha$ approaches $1$.
Further, we increase the rarity parameter $\gamma$ steadily so the probability of the rare-event becomes smaller. 
Again, the aim is to capture any numerical display of second order efficiency properties (c.f. Appendix~\ref{appendix:efficiency}) by the estimators. 

 We use the cross-entropy method for optimization~\cite[\S 13.2]{kroese2011handbook} to obtain the lower bound probability.
See appendix~\ref{appendix:matlab_ELM_scheme2} for the main program and functions used in {\tt MATLAB}.
We output the lower bound probability and estimated (rare-event) probability for each $\gamma$ and $\alpha$.
The relative error~\eqref{RE_formula} and relative time variance product~\eqref{RTVP_formula} for each of the estimators are given. 
Further, the total CPU time to obtain the ELM estimator is shown.

\paragraph{Parameters:}
In order to benchmark with ELM scheme A, we take similar algorithmic and problem parameters.
The dimension considered is $d=10$.
Sample sizes of $n_t=2\times 10^{4}$ for each density $f_t,t=1,2$. Total pooled sample size of $n=4\times 10^{4}$.
We use $K=100$ iterations to estimate the ELM relative error. 

\paragraph{Discussion of Results:}
The numerical results of the lower bound ELM scheme are given in Table~\ref{tab:ELS_scheme2} below. 
When $\alpha$ is close to zero, we see that the lower bound probabilities are very close to the (rare-event) probabilities. 
As $\alpha$ increase towards 1, the lower bound becomes less tight and the relative error is larger. 
This suggests the proposed ELM estimator performs slightly worse when the distribution is less heavy-tailed.
However, we observe a significant reduction in the overall relative error when comparing to the results of ELM scheme A.
Hence, the addition of the lower bound constraint seem to provide a better ELM estimator than the scheme with only equality constraints. 
We still observe numerical bounded relative error properties for this ELM scheme.

Compared to ELM scheme A, we observe higher CPU time to calculate the ELM estimator.
The bottleneck is due to solving the maximization problem for the optimal $\boldsymbol \lambda^*$ and lower bound probability $\ell_1$.
There seem to be some inconsistency in the optimization time in finding lower bound probability. Generally, it varies between $8$ to $10$ seconds observed for $\alpha=0.8$ and $0.9$.
However, the optimization takes significantly less time for $\alpha=0.1$ and large values of $\gamma$ for $\alpha=0.2$.
Instead using cross-entropy optimization, we might want to consider using a gradient optimization procedure in the future. 
This may decrease the overall computational time and further improve efficiency. 
Note, the overall RTVP for each $\gamma$ and $\alpha$ is still smaller for ELM scheme B than ELM scheme A.
This suggests that ELM scheme B, with the inclusion of lower bound, is superior to the ELM scheme  that only incorporates  equality constraints.

\begin{table}[H]
  \centering
  \caption{ELM Scheme with 2 densities and inequality constraints}
		\footnotesize
    \begin{tabular}{|r|rrrrr|}
    \hline
  $\alpha=0.9$ &       &  &       &       &  \\
    $\gamma$ & $\widehat{\ell}_1$ & $\widehat{\ell}_s$ & $\text{RE}$ & $\text{RTVP}$ & CPU Time (s) \\
    \hline
    $30$  & $1.23/10^{4}$ & $1.33/10^{4}$ & $2.10/10^{4}$ & $3.69/10^{7}$ & 8.37 \\
    $40$  & $5.58/10^{7}$ & $6.28/10^{7}$ & $2.95/10^{4}$ & $7.46/10^{7}$ & 8.56 \\
    $50$  & $1.92/10^{9}$ & $2.25/10^{9}$ & $3.84/10^{4}$ & $1.23/10^{6}$ & 8.35 \\
    $60$  & $5.72/10^{12}$ & $7.02/10^{12}$ & $5.58/10^{4}$ & $2.71/10^{6}$ & 8.70 \\
    $90$  & $1.10/10^{19}$ & $1.56/10^{19}$ & $1.07/10^{3}$ & $9.98/10^{6}$ & 8.76 \\
    \hline
    \hline
    $\alpha=0.8$      &       &  &       &       &  \\
    $\gamma$ & $\widehat{\ell}_1$ & $\widehat{\ell}_s$ & $\text{RE}$ & $\text{RTVP}$ & CPU Time (s) \\
    \hline
    $30$  & $1.39/10^{3}$ & $1.59/10^{3}$ & $2.31/10^{4}$ & $4.46/10^{7}$ & 8.37 \\
    $50$  & $7.82/10^{7}$ & $1.02/10^{6}$ & $5.06/10^{4}$ & $2.45/10^{6}$ & 9.60 \\
    $100$ & $8.65/10^{15}$ & $1.45/10^{14}$ & $2.40/10^{3}$ & $5.31/10^{5}$ & 9.18 \\
    $150$ & $9.23/10^{22}$ & $1.32/10^{21}$ & $3.24/10^{3}$ & $9.69/10^{5}$ & 9.26 \\
    $200$ & $3.96/10^{28}$ & $5.12/10^{28}$ & $1.80/10^{3}$ & $2.85/10^{5}$ & 8.81 \\
    \hline
    \hline
    $\alpha=0.6$      &       &  &       &       &  \\
    $\gamma$ & $\widehat{\ell}_1$ & $\widehat{\ell}_s$ & $\text{RE}$ & $\text{RTVP}$ & CPU Time (s) \\
    \hline
    $100$ & $6.62/10^{6}$ & $9.49/10^{6}$ & $1.00/10^{3}$ & $1.10/10^{5}$ & 10.97 \\
    $150$ & $5.92/10^{8}$ & $7.85/10^{8}$ & $1.23/10^{3}$ & $1.47/10^{5}$ & 9.71 \\
    $200$ & $1.09/10^{9}$ & $1.35/10^{9}$ & $8.19/10^{4}$ & $6.68/10^{6}$ & 9.95 \\
    $500$ & $1.67/10^{17}$ & $1.83/10^{17}$ & $3.51/10^{4}$ & $1.27/10^{6}$ & 10.31 \\
    $1000$ & $6.58/10^{27}$ & $7.00/10^{27}$ & $2.11/10^{4}$ & $4.42/10^{7}$ & 9.95 \\
    \hline
    \hline
    $\alpha=0.2$      &       &  &       &       &  \\
    $\gamma$ & $\widehat{\ell}_1$ & $\widehat{\ell}_s$ & $\text{RE}$ & $\text{RTVP}$ & CPU Time (s) \\
    \hline
    $10^{4}$ & $1.85/10^{2}$ & $1.97/10^{2}$ & $2.45/10^{4}$ & $5.25/10^{7}$ & 8.77\\
    $10^{6}$ & $4.56/10^{4}$ & $4.65/10^{4}$ & $1.22/10^{4}$ & $1.49/10^{7}$ & 10.01 \\
    $10^{6}$ & $1.31/10^{6}$ & $1.31/10^{6}$ & $4.90/10^{5}$ & $1.21/10^{8}$ & 5.04 \\
    $10^{7}$ & $1.23/10^{10}$ & $1.23/10^{10}$ & $1.79/10^{5}$ & $1.59/10^{9}$ & 4.96 \\
    $10^{8}$ & $5.13/10^{17}$ & $5.13/10^{17}$ & $6.50/10^{6}$ & $2.02/10^{10}$ & 4.79 \\
    \hline
    \hline
    $\alpha=0.1$ &       &  &       &       & \\
    $\gamma$ & $\widehat{\ell}_1$ & $\widehat{\ell}_s$ & $\text{RE}$ & $\text{RTVP}$ & CPU Time (s)\\
    \hline
    $10^{10}$ & $4.54/10^{4}$ & $4.55/10^{4}$ & $3.68/10^{5}$ & $4.70/10^{9}$ & 3.48\\
    $10^{11}$ & $3.40/10^{5}$ & $3.41/10^{5}$ & $2.89/10^{5}$ & $2.91/10^{9}$ & 3.50 \\
    $10^{12}$ & $1.31/10^{6}$ & $1.31/10^{6}$ & $2.67/10^{5}$ & $2.33/10^{9}$ & 3.26 \\
    $10^{13}$ & $2.13/10^{8}$ & $2.16/10^{8}$ & $8.68/10^{5}$ & $2.32/10^{8}$ & 3.09 \\
    $10^{15}$ & $1.85/10^{13}$ & $1.85/10^{13}$ & $3.73/10^{6}$ & $5.54/10^{11}$ & 3.99\\
    \hline
    \end{tabular}%
  \label{tab:ELS_scheme2}%
\end{table}%

\section{Critical Analysis of ELM Method}
In this section, we give a critical analysis of the proposed ELM method when applied to rare-event probability estimation.
The proposed ELM method has advantages in terms of versatility, general purpose, and usage of flexible constraints.
It formulates a convex optimization program which is typically easy to solve.   
Conversely, the ELM approach can be hindered by high computational costs. Dependent MCMC samples used may impact on the estimation.
Good reference densities can be difficult to find (but  are always easier to find than a good importance sampling density).


\paragraph{Strengths:}

\begin{enumerate}
\item Versatility and General Purpose --
In the ELM procedure, we can include the samples of any probability density that is known or known up to a multiplicative constant and use it directly in the estimation.  A major  advantage of ELM densities is that their choice is less restricted than the choice of valid importance sampling densities.
The only requirement is the connectivity condition on the supports of the given sequence of densities.
Of course, the density (or densities) should somehow be related to the integral we are trying to estimate. In this respect, the situation is similar to the use of control variables in Monte Carlo variance reduction --- the inclusion of any control variable reduces the variance, but if the control variable is not highly correlated to the quantity of interest, then the additional computational cost may not be justified. 

\item Flexibility of Constraints -- 
We can add in constraints in the optimization to account for any analytical knowledge of reference densities and their relationships. 
This would typically reduce the variance of the estimator,  again similar to the usage of control variates in Monte Carlo methods. 
For instance, in ELM scheme B, we have seen the inclusion of a lower bound reference density improved the efficiency of the ELM estimator.
In the future, it may be possible to find a reference density whose normalizing constant is the upper bound to the rare-event probability.
This would give a 100\% confidence interval for our estimator and its knowledge can be simply incorporated by using an inequality constraint.

\item Convex Optimization -- 
The ELM procedure results in the optimization of a convex (negative) log-likelihood function.
If the constraints added are linear, we will still have a convex optimization program. Convexity typically makes the optimization easier to solve.
For example, any local minimum solution found must also be a global minimum. 

\end{enumerate}

\paragraph{Shortcomings:} 


\begin{enumerate}

\item Computational Cost -- 
In the ELM method, a significant portion of the computational cost occurs from sampling from densities. 
The computational speed may be hindered by densities that are difficult to sample from. 
Typically, this is due nature of MCMC algorithms. 
For example, the Gibbs sampling of the probability density $f_3$ in ELM scheme A increased the average CPU time approximately 5.5 times.
Significant computational costs will increase RTVP which can severely impact efficiency .

\item Dependent Samples -- 
The theory of ELM relies on assuming iid samples. 
However, in practice, we need to employ MCMC methods to sample from the zero variance density and any density difficult to sample from.
Hence, the resulting samples are from a Markov chain which will have some degree of dependence. We are unsure of the impact on estimation.
One solution may be looking at the autocorrelations of the samples and discarding certain number of realizations to remove dependence. 
This is not ideal as it would significantly increase the computational cost. Alternatives like splitting and sequential Monte Carlo methods may be better and need to be explored. 

\item Availability of Densities -- For a specific problem, it is sometimes difficult to obtain an appropriate sequence of densities. 
Nevertheless, this is easier than finding a good importance sampling density. Of course, good ELM densities  are often inspired by, but not limited to, existing importance sampling methodologies.

\end{enumerate}

\chapter{Conclusion}

Empirical Likelihood Maximization (ELM) is a novel Monte Carlo method used in rare-event probability estimation.
The underlying mechanism involves the construction of an empirical likelihood using samples from a sequence of probability densities which may or may not  be sampled sequentially. 
These densities have normalizing constants which we treat as unknown parameters to be estimated. 
In particular, the (rare-event) probability of interest is embedded in the empirical likelihood via the zero-variance density.
Similar to likelihood maximization, we solve an optimization program to recover the estimator of the (rare-event) probability. A nice feature of the empirical likelihood is its convexity ensuring that we easily obtain a global solution to the optimization. 

There exist many Monte Carlo methods used in rare-event probability estimation which are specialized algorithms and would only perform well in certain scenarios while poorly in others.  
The proposed ELM method is very flexible and can be easily adapted to obtain an efficient estimator for a general problem.
Another advantage of the ELM method is its ability to utilize analytical knowledge of reference densities and their interrelationships.
The idea is similar to using a control variate -- a variance reduction technique used in Monte Carlo methods.
This is achieved by simply adding linear equality or inequality constraints in the convex optimization.  

Numerical experiments have been performed using this new technique and benchmarks are made against existing algorithms and estimators. 
The performance of the ELM approach has been consistently reliable and fares well against both the MCIS and AK conditional estimator. 
We have observed numerical bounded relative error properties for the ELM approach. 
However, one shortcoming of the ELM procedure is its computational speed to sample from difficult densities via MCMC.
When considering different ELM schemes, one needs to consider the trade-off between variance reduction and computational cost.
Some possible directions of future research are given below.

\paragraph{Future Research}

\begin{enumerate}

\item 
From our numerical experiments, we have observed the ELM estimator exhibits bounded relative error properties. 
We would like to make theoretical developments in establishing higher order efficiency properties. 
In particular, we want to establish the asymptotic behaviour of the ELM estimator as $\gamma\rightarrow\infty$. 

\item 
Currently, the relative error of the ELM estimate must be computed empirically from independent runs of the ELM scheme. 
If we had iid samples, we could use the Fisher information matrix corresponding to $\widehat{D}(\bz)$ to obtain an asymptotic result.
However, the dependence in MCMC sampling makes it difficult to obtain a central limit result for the empirical likelihood.
The current approach is inadequate and time consuming. 
We would like to develop an analytical formula for computing the relative error when we have linear constraints in the optimization and with MCMC sampling.
This may involve looking at the asymptotic behaviour as the events become rarer ($\gamma \rightarrow \infty$) and  when sample size becomes larger ($n\rightarrow\infty$). 

\item 
We would like to explore alternatives to MCMC sampling.
One possibility is the Sequential Monte Carlo (SMC) method.
As mentioned in section~\ref{splitting}, the SMC or splitting method does not yield competitive estimators compared to the AK estimator.
However,  using splitting or SMC only  for approximate sampling from the zero-variance density $\pi$ in~\eqref{eq:min_var_IS} may be advantageous.
Finally, we would like to explore the possibility of using upper bound inequality constraints. While we have found a method to obtain lower bounds, we are not presently able to find a variational approach that yields useful upper bounds (standard approaches like Chebyshev or Chernoff bounds are too loose to be useful).

\end{enumerate}


\begin{appendices}
\chapter{Miscellaneous Concepts}


\section{Efficiency}\label{appendix:efficiency}

The probability of interest $\ell$ usually depends on a rarity parameter $\gamma$ such that $\ell$ approaches zero as the threshold level $\gamma$ approaches infinity. That is,
$$\ell = \ell(\gamma) = \mathbb{P}(S(\bX)\geq \gamma) $$ where  $$\ell(\gamma)\downarrow 0 \quad \text{as } \quad \gamma \rightarrow \infty.$$
Thus, it is of interest to quantify the behaviour of an estimator as the event of interest becomes more rare. A key issue is whether the estimator's accuracy will deteriorate as the rare-event probability $\ell$ approaches zero. 
We introduce the following \emph{second order efficiency measures} from \cite[\S 10.1]{kroese2011handbook}.

\Definition 
Consider a random variable  $Z=Z(\gamma)$ with the above property. That is,  $$\Em(Z(\gamma))=\ell(\gamma)\rightarrow 0^+ \quad \text{as} \quad \gamma\rightarrow \infty.$$ 
\begin{enumerate}
\item The given estimator has (asymptotically) \emph{vanishing relative error} if 
$$ \limsup_{\gamma\rightarrow\infty}\frac{Var(Z(\gamma))}{\ell^2(\gamma)}=0.$$
Here, the relative error approaches zero as the event becomes more rare.
\item The given estimator has \emph{bounded relative error} if
$$ \limsup_{\gamma\rightarrow\infty}\frac{Var(Z(\gamma))}{\ell^2(\gamma)} \leq K < \infty,$$
where $K$ is a constant that does not depend on $\gamma$.
The relative error is bounded regardless of the rarity of the event. 
\item The given estimator is \emph{logarithmically efficient} if
$$ \limsup_{\gamma\rightarrow\infty}\frac{Var(Z(\gamma))}{\ell^{2-\epsilon}(\gamma)}=0$$  
for all $\epsilon > 0$, or equivalently if
$$ \liminf_{\gamma\rightarrow\infty}\left | \frac{\ln Var(Z(\gamma))}{\ln \ell^2(\gamma)}\right | \geq 1.$$
Here, the second moment of the estimator and the square of the its mean approach zero at the same exponential rate. 
\end{enumerate}
The vanishing relative error is a stronger condition than the bounded relative error. Both these conditions are stronger than logarithmic efficiency. An estimator with vanishing error property is the Asmussen-Kroese conditional estimator (see~\ref{AKCE}) for distributions with subexponential properties.  Generally, it is more difficult to find an estimator with bounded relative error than a logarithmically efficient estimator.  

\begin{Example}
{(Inefficiency of CMC Estimator)}
\label{appendix:CMC}
The naive Crude Monte Carlo (CMC) estimator is given by $$Z = Z(\gamma) = \I\{S(\bX) \geq \gamma \}.$$ 
For this estimator $$\Em Z^2 = \Em Z = \ell$$ and thus $$\lim_{\gamma \rightarrow \infty}\frac{\log \Em Z^2}{\log \ell^2}=\frac{\log \ell}{\log \ell^2} = \frac{1}{2}<1.$$ 
Hence, the CMC estimator is not logarithmically efficient. For small $\ell$, the relative error $\epsilon$ of the CMC estimator satisfies
$$\epsilon = \frac{\sqrt{\text{Var}(\widehat{\ell})}}{\Em \widehat{\ell}}=\frac{\sqrt{\frac{1}{N}\text{Var}{Z}}}{\ell}=\sqrt{\frac{1-\ell}{N\ell}}\approx \sqrt{\frac{1}{N\ell}}.$$
Suppose $\ell=10^{-6}$. Then to estimate $\ell$ accurately with relative error of $1\%$, we require a sample size of 
$$N \approx \frac{1}{\epsilon^2 \ell}=10^{10}.$$
The computational cost is too high and using the CMC estimator becomes an infeasible proposition for rare-events. Thus, this motivates the need for improved methods and alternate estimators of rare-event probabilities. 

\end{Example}

We briefly note there exist potential problems with the second-order efficiency measures as they do not convey any information about the accuracy of the estimator $\widehat{\sigma}^2$ of the true variance $\sigma^2$. This issue is addressed by the introduction of higher-order efficiency measures as outlined in~\cite[\S 10.1]{kroese2011handbook} and \cite{l2010asymptotic}.

\section{Tail properties of distributions}
\label{appendix:tail_properties}
A random variable $X$ with cdf $F$ is said to have a (right) \emph{light-tailed} distribution if its moment generating function is finite for some $t>0$. That is,
\begin{equation}
\label{eq:light-tail}
\Em(e^{tX}) \leq c < \infty.
\end{equation}
Otherwise, when $\Em(e^{tX}) =\infty$ for all $t>0$, $X$ is said to have \emph{heavy-tailed} distribution. 
We only consider the right tail of the distribution as the left tail can be treated equivalently. 
Note for every $x$
$$ \Em e^{tX} \geq \Em e^{tX} \I(X>x) \geq e^{tx} \mathbb{P}(X>x).$$ 
It follows that for any light-tailed distribution satisfying~\eqref{eq:light-tail}
$$\mathbb{P}(X>x)\leq ce^{-tx}.$$
That is, 	if $X$ has a light-tailed distribution, then $\bar{F}(x) = 1-F(x)$ decays at an exponential rate or faster. Similarly, heavy-tailedness property is equivalent to 
$\lim_{x\rightarrow\infty} e^{tx} \bar{F}(x)  = \infty$ for all $t>0$.
On the other hand, any distribution with bounded support is light-tailed. Examples of light-tailed distributions with unbounded support include Exponential, Geometric, Poisson, Gamma, and \textsf{Weibull}($\alpha,\lambda$) for $\alpha \geq1$.
Further, there are additional properties that imply heavy-tailedness. They are given below in the following order of generality:
$$\text{Regularly varying} \Rightarrow \text{Subexponential} \Rightarrow \text{Long-tailed} \Rightarrow \text{Heavy-tailed}.$$

\begin{enumerate}
\item A distribution is said to be \emph{long-tailed} if
$$\lim_{x\rightarrow\infty} \frac{\bar{F}(x+t)}{\bar{F}(x)}=1, \quad \text{for all} \ t.$$
\item A distribution on the interval $(0,\infty)$ is said to be \emph{subexponential} if, with $X_1,\dots,X_n \stackrel{iid}{\sim} F$
$$\lim_{x\rightarrow\infty}\frac{\mathbb{P}(X_1+\cdots+X_n > x)}{\mathbb{P}(X_1 > x)}=n, \quad \text{for all} \ n,$$
or equivalently,
$$\lim_{x\rightarrow\infty}\frac{\mathbb{P}(X_1+\cdots+X_n > x)}{\mathbb{P}(\max\{X_1,\dots,X_n\} > x)}=1.$$
\item A distribution is called \emph{regularly varying} if
$$\bar{F}(x)=\frac{L(x)}{x^{\alpha}} $$
for some $\alpha >0$ and some function $L$ that satisfies $L(tx)/L(x)\rightarrow 1$ as $x\rightarrow\infty$ for all $t>0.$
\end{enumerate}
Examples of regularly varying distributions include Student-t, Cauchy, and Pareto.
Two families of distributions that are subexponential but not regularly varying are Log-Normal and  \textsf{Weib}($\alpha,\lambda$) for $\alpha <1$. 
For more detailed class properties of heavy-tailed distributions see~\cite{embrechts1982estimates}.

\section{Distance Measures} \label{appendix:distance_measures}

Instead of Kullback-Leibler distance, there exist other distance or divergence measures between probability density functions. An important class of distance measures used to capture the ``closeness'' between two pdfs $f$ and $g$, is the \emph{Csisz\'{a}r's $\phi$-divergence}
\begin{equation}\label{eq:phi_divergence}
d(f,g)=\int f(\bx) \phi \left( \frac{g(\bx)}{f(\bx)} \right) \ d\bx, 
\end{equation}
where $\phi:\mathbb{R}^+ \rightarrow \mathbb{R}$ is twice continuously differentiable with $\phi(1)=0$ and $\phi''(x) > 0$ for all $x>0$. Note that $\phi$ is a convex function.
Special cases~\cite{rubinstein2011simulation} of the $\phi$--divergence include most of the information-theoretic distances such as Burg CE distance,  Kullback-Leibler CE distance, Hellinger distance and Pearson $\chi^2$ distance. They are given below
\begin{itemize}
\item Burg CE distance:
$$ d(f,g)=\int g(\bx)\log \frac{g(\bx)}{f(\bx)} \ d\bx.$$

\item Kullback-Leibler CE distance:
$$ d(f,g)=\int f(\bx)\log \frac{f(\bx)}{g(\bx)} \ d\bx.$$

\item Hellinger distance:
$$ d(f,g)=2\int \left ( \sqrt{f(\bx)}-\sqrt{g(\bx)}\right)^2 \ d\bx.$$

\item Pearson $\chi^2$ discrepancy measure:
$$ d(f,g)=\frac{1}{2}\int \frac{[f(\bx)-g(\bx)]^2}{g(\bx)} \ d\bx.$$
\end{itemize}
Formally, $d(f,g)$ is not a distance between $f$ and $g$ as $d(f,g) \neq d(g,f)$ in general. However, it is useful to think of it as a distance measure because $$d(f,g) \geq 0$$ and $d(f,g)=0$ if and only if $g(\bx)=h(\bx)$. This follows from Jensen's inequality and convexity of $\phi$. Namely,
$$D(f,g)=\Em_f \left [ \phi\left(\frac{g(\bX)}{f(\bX)}\right)\right ] \geq \phi \left ( \Em_f \left [ \frac{g(\bX)}{f(\bX)}\right ]\right) = \phi(1) = 0.$$

\section{Random Variable Simulation} \label{random_variable_simulation}
Estimation and optimization problems in applied probability and statistics often concern sampling from specified target distributions.
Standard sampling methods include the inverse-transform method, the acceptance-rejection method. These methods are exact, that is, the generated random variables are distributed exactly according to the target distribution.

\subsection{Inverse Transform Method}
Let $X$ be a random variable with cdf $F$. Since $F$ is a non-decreasing function, the inverse cdf $F^{-1}$ can be defined as
$$\{F^{-1}(y)= \inf\{x: F(x)\geq y\} \}, \quad 0 \leq y \leq 1. $$
Let $U\sim \textsf{Uniform}(0,1)$. The cdf of the inverse transform $F^{-1}(U)$ is given by
$$ \mathbb{P}(F^{-1}(U) \geq x) = \mathbb{P}(U \leq F(x))= F(x).$$
Hence, to simulate a random variable from $X$ with cdf $F$, draw $U\sim \textsf{Uniform}(0,1)$ and set $X=F^{-1}(U).$

\begin{Algorithm} \label{algo:Inverse-Transform}
(Inverse-Transform Method)
\begin{enumerate}
\item Generate $U\sim \textsf{Uniform}(0,1)$.
\item Deliver $X=F^{-1}(U).$
\end{enumerate}
\end{Algorithm}
It is sometimes of interest to apply the inverse transform method to sample from a truncated probability density. More detail is given below.

\paragraph{Truncation:} Suppose $G_{\mathcal{A}}$ and $G_{\mathcal{B}}$ are two distributions on sets $\mathcal{A}$ and $\mathcal{B}\subset\mathcal{A}$, respectively. Let
$$\bX \sim G_{\mathcal{A}}, \quad \bZ \sim G_{\mathcal{B}}.$$ 
If the conditional distribution of $\bX$ given $\bX\in\mathcal{B}$ coincides with the distribution of $\bZ$, then the latter distribution is said to be the truncation of $G_{\mathcal{A}}$ to $\mathcal{B}$. For this case if $f_{\bX}$ is the pdf of $\bX$, then the pdf of $\bZ$ is $$f_{\bZ}(\bz) = \frac{f_{\bX}(\bz)}{\int_{\mathcal{B}}f_{\bX}(\bx)\ d\bx}, \quad \bz \in \mathcal{B}.$$
In the continuous univariate case, the truncation of a density $f(x)$ to an interval $[a,b]$ gives the pdf
$$ f_Z(z) = \frac{f(z)}{\int_{a}^b f(x) \ dx}, \quad a\leq x\leq b.$$
Writing in terms of cdfs, we have
$$F_Z{(z)} = \frac{F(z)-F(a^{-})}{F(b)-F(a^{-})}, \quad a\leq x\leq b,$$
where $F(a^-) = \lim_{x\uparrow a}F(x)$. When the generation of $X$ can be readily performed via the inverse-transform method, we can simulate from the truncated distribution using algorithm~\reff{random_variable_simulation}{algo:truncation} below.

\begin{Algorithm}\label{algo:truncation}
(Truncation with the Inverse-Transform Method)
\begin{enumerate}
\item Generate $U\sim \textsf{Uniform}(0,1)$.
\item Deliver $Z=F^{-1}\left(F(a^-)+U(F(b)-F(a^-))\right).$
\end{enumerate}
\end{Algorithm}
The only difference with the inverse-transform algorithm~\reff{random_variable_simulation}{algo:Inverse-Transform} is in step 2 where the argument of $F^{-1}$ is uniformly distributed on the interval $(F(a-),F(b))$ rather than on $(0,1)$.


\subsection{Acceptance-Rejection Method}
The inverse-transform method will fail when the inverse of the cdf does not have a closed form.
An alternative exact method of sampling is the acceptance-rejection method. It is one of the most useful methods for sampling from general distributions.
The method is based on the following observation.

\begin{Theorem}
{(Acceptance-Rejection)}
Let $f(\bx)$ and $g(\bx)$ be two pdfs such that for some $C\geq 1$, $Cg(\bx)\geq f(\bx)$ for all $\bx$. Let $\bX\sim g(\bx)$ and $U \sim \textsf{Uniform}(0,1)$ be independent.
Then, the conditional pdf of $\bX$ given $U \leq f(\bx) / (Cg(\bx))$ is $f(\bx)$.
\end{Theorem}

\noindent The proof is given chapter 3 of~\cite{kroese2011handbook}. The density $g(x)$ is called the proposal pdf and we assume that it is easy to generate random variables from it.
The acceptance-rejection method can be formulated below in algorithm

\begin{Algorithm}
(Acceptance-Rejection)
\begin{enumerate}
\item Generate $\bX$ from $g(\bx)$.
\item Generate $U$ from $\textsf{Uniform}(0,1)$ independently of $\bX$.
\item If $U \leq f(\bX) / (C g(\bX))$, output $\bX$. Else, reject $\bX$ and return to step 1.
\end{enumerate}
\end{Algorithm}
We generate $\bX \sim g$ and accept it with probability $ f(\bX) / (C g(\bX))$, otherwise reject $\bX$ and try again.

\section{Markov chain Monte Carlo}
\label{sec:MCMC}

Markov chain Monte Carlo (MCMC) is a generic method used for approximate sampling from an arbitrary distribution.
The key idea is to generate a Markov chain whose limiting distribution is equal to the desired distribution.    
It is utilized in many cases where exact sampling is very costly or infeasible.

The idea of MCMC was first introduced by Metropolis, Rosenbluth, Rosenbluth, Teller and Teller (1953) in~\cite{metropolis1953equation} as a method for efficient simulation of the energy levels of atoms in a crystalline structure. It is subsequently adapted by Hastings (1970) in~\cite{hastings1970monte} to focus upon statistical problems and is now frequently used in Bayesian statistics.  

Here we focus two prominent MCMC algorithms: the \emph{Metropolis-Hastings} algorithm and, its special case, the \emph{Gibbs Sampler}. This algorithm has been applied several times during our numerical experiments in Chapter 4. 

\subsection{Metropolis-Hastings Algorithm}
This is a general way to construct the MCMC sampler. It is analogous to rejection sampling where realizations are drawn from approximate distributions. However, instead of rejecting samples, we correct them so they asymptotically behave as random observations from the target distribution.
The method induces a Markov chain by sequentially drawing candidate observations from a distribution conditional on the last observation.

\paragraph{Formulation:} Suppose that we want to generate samples from an arbitrary  multi-dimensional probability density
$$ f({\bf{x}}) = \frac{p({\bf x})}{\mathcal{Z}}, \quad {\bf{x}} \in \mathcal{X} \subseteq \mathbb{R}^n, $$
where $p({\bf x})$ is a known positive function, and $\mathcal{Z}$ is a known or unknown normalizing constant.
Let $q({\bf y} | {\bf x}) $ be the \emph{proposal density} or \emph{candidate generating function}. This is a Markov transition density describing how to go from state ${\bf x}$ to ${\bf y}$. When a process is at the point ${\bf x}$, the density generates a value ${\bf y }$ from $q({\bf y} | {\bf x}) $.
Algorithm~\reff{sec:MCMC}{algo:MHA} below illustrates how a Markov chain with state space $\mathcal{X}$ and stationary distribution $f({\bf{x}})$ is constructed.

\begin{Algorithm} (Metropolis-Hastings Algorithm)
\label{algo:MHA}
To sample from a density $f(\bx$) known up to a normalizing constant, initialize with some $\bX_0$ for which $f({\bX_0}) > 0.$
For each $t = 0, 1,2, \dots,T-1$, execute:
		\begin{enumerate}
		\item Given current state ${\bf X}_t$, generate ${\bf Y} \sim q({\bf y} | {\bf X}_t)$
		\item Generate $U \sim \mbox{Uniform}(0,1)$ and deliver
		$$ {\bf X}_{t+1} = \begin{cases}
		{\bf Y} & \mbox{if} \ U \leq \alpha ({\bX_t, \bY})\\
		{\bf X}_t & \mbox{otherwise},
		\end{cases}$$
		where
		$$ \alpha ({\bf x, y}) = \min \left \{ \frac{f({\bf y})q({\bf x | y})}{f({\bf x})q({\bf y|x})}, 1 \right \} .$$
		
		\end{enumerate}
\end{Algorithm}
The probability $ \alpha ({\bf x, y})  $ is the \emph{acceptance probability}. If the move is accepted, the process moves to ${\bf Y}$, otherwise it
remains at $\bX_t$. We can also replace $f$ by $p$ in the acceptance probability.

The resulting Metropolis-Hastings Markov chain is $\bX_0, \bX_1, \dots, \bX_T$ with $\bX_T$ being approximately distributed as $f({\bf x})$ for large $T$. In other words, if we run the chain for long enough, the realizations from the chain can be treated as a sample from the target distribution. However, the samples obtained would not be independent.
One Metropolis-Hastings iteration is equivalent to generating a point from the transition density $\kappa(\bx_{t+1}|\bx_t )$, where
$$\kappa({\bf y|x }) = \alpha ({\bf x, y}) q({\bf y|x}) + (1-\alpha^*({\bf x})) \delta_{\bf x} ({\bf y})$$
with
$$\alpha^*({\bf x}) = \int \alpha ({\bf x, y}) q({\bf y|x}) d{\bf y} \quad \mbox{and} \quad \delta_{\bf x} ({\bf y}) \ \mbox{as the Dirac delta function}.$$
The transition density is required to satisfy the detailed balance equation or also known as the reversibility condition
$$f({\bf x})\kappa({\bf y|x }) = f({\bf y})\kappa({\bf x|y }).$$
Hence, the density $f$ must be the stationary probability density of the Markov chain. 
Now, suppose the event $\{ \bX_{t+1} = \bX_t\}$ has positive probability. That is, the transition density $q$ satisfies the condition $$\mathbb{P} (\alpha({\bX_t, \bY})<1 |{\bX_t}) >0$$ and $$q({\by|\bx})> 0$$ for all ${\bf x,y \in \mathcal{X}}$. Then $f$ must also be the limiting density of the Markov chain.

With exception of trivial cases, as long as the chain has a probability of eventually reaching any state from any other state, the MCMC algorithm will work.
For an ergodic and irreducible Markov chain, the stationary distribution is also the limiting distribution of successive iterations from the chain.

\subsection{Gibbs Sampler}
\label{appendix_gibbs}
The Gibbs Sampler is a special case of the Metropolis-Hastings algorithm. The proposal or candidate generating function is just the conditional distribution $$q({\bf x|y}) = f({\bf x|y}).$$
Further, the acceptance probability is 1. In other words, every proposal ${\bf y}$ is accepted with $$ \alpha({\bf x,y}) = 1 \quad \forall ({\bf x,y }). $$
A distinguishing feature that sets apart the Gibbs Sampler is that the underlying Markov chain is built by a sequence of conditional distributions.
It is used widely whenever it is easy to sample from the conditional distributions of the joint density.

\paragraph{Formulation:} Suppose that we want to sample a random vector $ {\bf X} = (X_1,\dots,X_n)$ from a target probability density $f({\bf x})$.
Using the Bayesian notation, let $$f(x_i | \bx_{-i})=f(x_i | x_1,\dots,x_{i-1},x_{x+1},\dots,x_n)$$ represent the conditional probability density of the $i$-th component, $X_i$, given all the other components  except $x_i$. Algorithm~\reff{sec:MCMC}{algo:Gibbs} below describes the implementation of the Gibbs sampler.

\begin{Algorithm}
\label{algo:Gibbs}
 (Gibbs Sampler)
 Given an initial state ${\bf X}_0$, iterate the following steps for $t=0,1,\dots$
		\begin{enumerate}
		\item For a given ${\bf X}_t$, generate $ {\bf Y } = (Y_1,\dots,Y_n)$ as follows:
			\begin{enumerate}
			\item Draw $Y_1$ from $f(x_1|X_{t,2}, \dots, X_{t,n})$.
			\item Draw $Y_i$ from $f(x_i|Y_1,\dots,Y_{i-1},X_{t,i+1}, \dots, X_{t,n})$ for $i =2,\dots, n-1$.
			\item Draw $Y_n$ from $f(x_n|Y_1,\dots,Y_{n-1})$.
			\end{enumerate}
		\item Set ${\bf X}_{t+1} = {\bf Y}$.
		
		\end{enumerate}
\end{Algorithm}
The transition probability density function is given by $$\kappa_{1\rightarrow n} ({\bf y | x}) = \prod_{i=1}^n{f(y_i | y_1,\dots,y_{i-1},x_{i+1},\dots,x_n)}, $$
where the subscript $1\rightarrow n$ indicates that the components of vector ${\bf x}$ are updated in the systematic order $1\rightarrow 2\rightarrow 3 \rightarrow \cdots \rightarrow n$.
Now if the transition density of the reverse move $\by \rightarrow \bx$, then the vector ${\bf y}$ is updated in order $n\rightarrow n-1\rightarrow \cdots \rightarrow 1$. In this case, we have
$$\kappa_{n\rightarrow 1} ({\bf x | y}) = \prod_{i=1}^n{f(x_i | y_1,\dots,y_{i-1},x_{i+1},\dots,x_n)}.$$
Algorithm~\reff{sec:MCMC}{algo:Gibbs} is known as the systematic or coordinate-wise Gibbs Sampler.
The completion of all the conditional sampling steps in the specified order is called a cycle.

There are alternative ways to update the components of $\bX$. 
In the random sweep or scan Gibbs sampler, a single cycle can consist of one or several coordinates selected uniformly from the integers $1,\dots,n$ or a random permutation 
$\pi_1 \rightarrow \pi_2 \rightarrow \cdots \rightarrow \pi_n$ of all coordinates. The resulting Markov chain $\{\bX_t, t=1,2,\dots \}$ is reversible. 
In the situation where a cycle consists of a single randomly selected coordinate, the random Gibbs sampler can be viewed as a Metropolis-Hastings sampler with transition function
$$q(\by|\bx) = \frac{1}{n} f(y_i|\bx_{-i}) = \frac{1}{n} \frac{f(\by)}{\sum_{y_i} f(\by)}, $$
where $\by = (x_1,\dots,x_{i-1}, y_i, x_{i+1},\dots,x_n)$. Since $\sum_{y_i} f(\by)$ can 	also be written as $\sum_{x_i} f(\by)$, we have
$$\frac{f(\by)q(\bx|\by)}{f(\bx)q(\by|\bx)} = \frac{f(\by)f(\bx)}{f(\bx)f(\by)}=1, $$
so in this case,  the acceptance probability $\alpha(\bx,\by)$ is 1. 

%

\end{appendices}


\begin{appendices}
\chapter{Miscellaneous Proofs}

\section{Lemma~\reff{MCIS}{lemma:POM}} \label{appendix:lemma_POM}
Suppose that $\pi_i(y_i)$ is the marginal of the zero-variance density $\pi(\by)$. Then, the solution to functional optimization program~\eqref{eq:FOP} is 
$$g_i(y_i)=\pi_i(y_i) \quad \text{for all} \quad i=1,\dots,d.$$
That is, the best importance sampling density (in the cross-entropy sense) within the space of all product-form densities is the product of marginals for~\eqref{eq:zero_variance_density}.
This is given by
\begin{align}
\label{eq:POMs2}
g(\by) = \prod_{i=1}^d\pi_i(y_i).
\end{align}

\begin{proof} Consider
\begin{align*}
d(\pi,g) & = \int \pi(\by) \log \left( \frac{\pi(\by)}{\prod_{i=1}^d g_i(y_i)}\right) \di\by\\
& = \int \pi(\by) \log \left(\pi(\by)\right)\di\by - \int \pi(\by) \sum_{i=1}^d\log (g_i(y_i)) \di\by\\
& = \int \pi(\by) \log \left(\pi(\by)\right)\di\by - \sum_{i=1}^d \int \pi_i(y_i) \log (g_i(y_i)) \di y_i\\
& = \sum_{i=1}^d \int \pi_i(y_i)\log\left(\frac{\pi_i(y_i)}{g_i(y_i)}\right) \di y_i + K(\pi(\by)),
\end{align*}
where $K(\pi(\by))$ consists of terms without $g_i$ for all $i=1,\dots,d$.
Hence, the functional optimization program~\eqref{eq:FOP} is equivalent to
\begin{align}
\min_{g_i \in \mathcal{G}} \int \pi_i(y_i) \log \left ( \frac{\pi_i(y_i)}{g_i(y_i)}\right) \di y_i, \quad \text{for all} \quad i=1,\dots,d. 
\end{align}
We recognize this as the cross-entropy distances between the marginals $\pi_i$ and $g_i$ for all $i$. These are zero if and only if $g_i = \pi_i$ for all $i$.

\end{proof}

\section{Lemma~\reff{ELS_Scheme1}{lemma:f2_simplify}}
\label{appendix:proof_f2_simplify}
The $i$th marginal from the zero-variance density can be written as 
\begin{equation}
\widehat{\pi}_i (x_i) = \frac{f(x_i)}{n_s} \sum_{j=1}^{n_s} e^{C_{ji}^{\alpha}} \I \{x_i \geq C_{ji} \},
\end{equation}
where $$C_{ji}= \max\left\{0,\gamma-\sum_{k\neq i} X_{jk}\right\}, $$
are calculated constants from samples of $f_s$.
It follows that
\begin{equation}
\frac{f_2(\bx)}{f(\bx)}= \prod_{i=1}^d \frac{1}{n_s}\sum_{j=1}^{n_s}  e^{C_{ji}^{\alpha}} \I \{x_i \geq C_{ji}\}, \ \bx=(x_1,\dots,x_d). 
\end{equation}

\begin{proof} Recall that our product of marginals density is
$$ f_2(\bx) = \prod_{i=1}^d \widehat{\pi}_i(x_i), \quad \bx = (x_1,\dots,x_d), $$
where $\widehat{\pi}_i(x_i)$ are the marginal densities of $f_s$.
Using samples $\bX_1, \dots , \bX_{n_s}$ generated from $f_s$, we can estimate $\widehat{\pi}_i(x_i)$ by their conditionals
$$\widehat{\pi}_i(x_i) = \frac{1}{n_s} \sum_{j=1}^{n_s} f_s(x_i | \bX_{j,-i}), $$
where $ \bX_{j,-i} = (X_{j,1},\dots, X_{j,i-1},X_{j,i+1},\dots, X_{j,d})$ is the $j$-th sample without the $i$-th element.
Now, by iid property of each sample
\[
f_s(x_i | \bX_{j,-i}) = 
\begin{cases}
f(x_i), & \mbox{if} \ \sum_{k\neq i} X_{jk} >\gamma \\
\\
f(x_i)\I\left\{ x_i \geq \gamma - \sum_{k\neq i} X_{jk} \right\}/K_{ji}, & \mbox{if} \  \sum_{k\neq i} X_{jk} \leq \gamma,
\end{cases}
\]
where $$K_{ji} = \Em_{f} \left( \I\left\{X \geq \gamma - \sum_{k \neq i} X_{jk}\right \}\right)=\mathbb{P}\left(X \geq \gamma - \sum_{k \neq i} X_{jk}\right) = \exp\left\{ -\left(\gamma - \sum_{k\neq i} X_{jk}\right)^\alpha\right\}$$ is the normalizing probability that a $\textsf{Weib}(\alpha,1)$ random variable exceeds some threshold. 
Now let us define
$$C_{ji}= \max\left\{0,\gamma-\sum_{k\neq i} X_{jk}\right\}. $$
Thus, we have
\begin{align*}
\frac{f_s(x_i | \bX_{j,-i})}{f(x_i)} 
& =  \I\left \{ \sum_{k\neq i}X_{jk} > \gamma\right \} + \frac{1}{K_{ji}}\I\left \{\sum_{k\neq i} X_{jk} \leq \gamma \right \} \I\left \{x_i \geq \gamma - \sum_{k\neq i} X_{jk} \right \} \\
& = \begin{cases}
\displaystyle 1 &  \text{if} \ \ \sum_{k\neq i}X_{jk} >\gamma \ \\ 
0 &\text{if} \ \ \sum_{k\neq i}X_{jk} \leq \gamma \ \ \text{and} \ \ x_i < \gamma - \sum_{k\neq i}X_{jk} \\
\displaystyle \frac{1}{K_{ji}} & \text{if} \ \ \sum_{k\neq i}X_{jk}  \leq \gamma \ \ \text{and} \ \ x_i \geq \gamma - \sum_{k\neq i}X_{jk}
\end{cases} \\
& =  \ e^{C_{ji}^{\alpha}} \ \I \left \{  x_i \geq C_{ji} \right\}. 
\end{align*}
Hence,
$$\widehat{\pi}_i (x_i) = \frac{f(x_i)}{n_s} \sum_{j=1}^{n_s} e^{C_{ji}^{\alpha}} \I \{x_i \geq C_{ji} \}, $$
which implies
$$ \frac{f_2(\bx)}{f(\bx)}= \prod_{i=1}^d \frac{1}{n_s}\sum_{j=1}^{n_s}  e^{C_{ji}^{\alpha}} \I \{x_i \geq C_{ji} \}.$$ 

\end{proof}

%
%
%
%

\section{Theorem~\reff{ELM_scheme_B}{lower_bound_lemma}} \label{lower_bound_lemma_proof}

For any positive value of the variational parameters $\lambda_1,\ldots,\lambda_d$, we have a lower bound to $\ell$  given by
\begin{equation}
\ell_{L} =  \Pm(S_L(\bY;\boldsymbol \lambda)\geq\gamma) = Q\left(\boldsymbol\beta;\alpha\gamma+(1-\alpha)\sum_i\lambda_i^{1/\alpha}\right),
\end{equation}
where
\begin{equation}
S_L(\by;\boldsymbol \lambda) \stackrel{def}{=} \frac{1}{\alpha}\sum_i \lambda_i^{1/\alpha-1} y_i-\frac{1-\alpha}{\alpha}\sum_i \lambda_i^{1/\alpha}
\end{equation} 
and the function $Q(\boldsymbol\beta;\gamma)=(1,0,\ldots,0) \;\e^{A\,\gamma}\; \mathbf{1}$ is defined via the matrix exponential with
\[
A=\left(\begin{array}{ccccc}
-\beta_1 & \beta_1 & 0 & \cdots &0 \\
0 & -\beta_2 & \beta_2 & \cdots &0 \\
\vdots & \vdots & \ddots & \ddots &\vdots \\
0 & \cdots & 0 & -\beta_{d-1} &\beta_{d-1}  \\
0 & \cdots & 0 & 0 &-\beta_{d}  \\
 \end{array}\right).
\]
and 
$$\beta_j=\frac{\lambda_j^{1/\alpha-1}+\cdots+\lambda_d^{1/\alpha-1}}{d-j+1}, \qquad j=1,\dots,d.$$

\begin{proof}

Consider the convex function $S(\by)=y_1^{1/\alpha}+\cdots+y_d^{1/\alpha}$. For convex functions, we have the following inequality
\[S(\by) \geq  S(\boldsymbol \lambda) + \nabla S(\boldsymbol \lambda)^{\top} (\by - \boldsymbol \lambda), \quad \text{for all valid}  \ \  \boldsymbol \lambda.\]
Hence,
\[
S(\by)\geq \frac{1}{\alpha}\sum_i \lambda_i^{1/\alpha-1} y_i-\frac{1-\alpha}{\alpha}\sum_i \lambda_i^{1/\alpha}\stackrel{def}{=} S_L(\by;\boldsymbol \lambda)
\]
is valid for all values of $\boldsymbol\lambda \geq \boldsymbol 0$. Thus, we have $\{S_L(\bY;\boldsymbol \lambda) \geq\gamma\}\subseteq\{S(\bY)\geq\gamma\} $.
It follows that
\[
\ell_L = \sup_{\boldsymbol \lambda \geq \boldsymbol 0} \ \Pm(S_L(\bY;\boldsymbol \lambda)\geq\gamma)\leq \Pm(S(\bY)\geq\gamma) = \ell.
\]
We now exploit the following representation of the order statistic $Y_{[1]},\ldots,Y_{[d]}$ of $\mathsf{Exp}(1)$ exponential random variables (see~\cite{devroye1986nonuniform}):
\[
Y_{[i]} = \sum_{j=1}^i \frac{Z_j}{d-j+1},\qquad Z_1,\ldots,Z_d\simiid \mathsf{Exp}(1).
\]
Note that
\[
\begin{split}
\sum_i Y_{[i]}\lambda_i^{1/\alpha-1}&=\sum_i \lambda_i^{1/\alpha-1} \sum_{j=1}^i \frac{Z_j}{d-j+1}\\
&= \sum_{j=1}^d Z_j \beta_j,\qquad \beta_j=\frac{\lambda_j^{1/\alpha-1}+\cdots+\lambda_d^{1/\alpha-1}}{d-j+1}
\end{split}
\]
Therefore, we have
\begin{equation*}
\begin{split}
\Pm(S_L(\bY;\boldsymbol \lambda)\geq\gamma)
&=\Pm\left(Y_{[1]} \lambda_1^{1/\alpha-1}+\cdots+Y_{[d]} \lambda_d^{1/\alpha-1}\geq\alpha\gamma+(1-\alpha)\sum_i\lambda_i^{1/\alpha}\right)\\
& =\Pm\left(\sum_{j=1}^d Z_j\beta_j \geq \alpha\gamma+(1-\alpha)\sum_i\lambda_i^{1/\alpha}\right) \\
&=Q\left(\boldsymbol\beta;\alpha\gamma+(1-\alpha)\sum_i\lambda_i^{1/\alpha}\right), \\
\end{split}
\end{equation*}
where the function $Q(\boldsymbol\beta;\gamma)=(1,0,\ldots,0) \;\e^{A\,\gamma}\; \mathbf{1}$ is defined via the matrix exponential with
\[
A=\left(\begin{array}{ccccc}
-\beta_1 & \beta_1 & 0 & \cdots &0 \\
0 & -\beta_2 & \beta_2 & \cdots &0 \\
\vdots & \vdots & \ddots & \ddots &\vdots \\
0 & \cdots & 0 & -\beta_{d-1} &\beta_{d-1}  \\
0 & \cdots & 0 & 0 &-\beta_{d}  \\
 \end{array}\right).
\]
Note $Q$ is the tail probability of a Generalized Erlang distribution. This is a specific case of a phase-type distribution. For more details, see~\cite[\S 4.2.13]{kroese2011handbook}.

\end{proof}

\end{appendices}

\begin{appendices}
\chapter{{\tt MATLAB} Code}

This appendix chapter will contain the main {\tt MATLAB} programs and functions used in the numerical examples from sections~\ref{numerical_results1} and~\ref{numerical_results2}.

\section{ELM scheme A: 4 densities and equality constraints}
\label{appendix:matlab_ELM_scheme1}

The main program {\tt weibull\_main\_A.m} computes the estimator from the ELM scheme A.
Also, the MCIS estimator and AK-conditional estimator are computed along with their relative errors.
This main program calls functions used to sample from the 4 densities and functions used for the constrained empirical likelihood optimization.
This is illustrated in figure~\ref{fig:functions_ELM1} below.

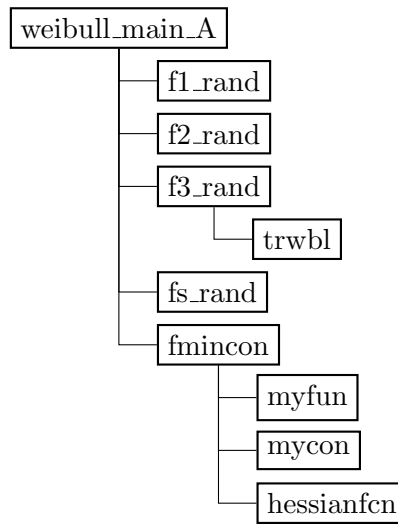
\begin{figure}[H]
\centering
\caption{Functions for ELM Scheme A} \label{fig:functions_ELM1}
\begin{tikzpicture}[%
  grow via three points={one child at (0.5,-0.7) and
  two children at (0.5,-0.7) and (0.5,-1.4)},
  edge from parent path={(\tikzparentnode.south) |- (\tikzchildnode.west)}]
  \node { weibull\_main\_A}
    child { node {f1\_rand}}		
    child { node {f2\_rand}}
    child { node {f3\_rand}
      		child { node {trwbl}}
    }
    child [missing] {}	
    child { node {fs\_rand}}
    child { node {fmincon}
      child { node {myfun}}
      child { node {mycon}}
      child { node {hessianfcn}}
    };
\end{tikzpicture}
\end{figure}

\subsection{Main program}

\begin{Verbatim}[fontsize=\footnotesize,frame=topline,label=Main program {\tt weibull\_main\_A.m}]

%%%
% weibull_main_A.m                                                
% Date: 23/10/2013                                                                        
% Description: Main program to compute estimators from ELM scheme A, MCIS
% and AK procedure
%%%

clear all, clc
format long g

%% Problem Parameters
d=10;
alpha=0.9; 
gamma=90;

%% ELM Scheme 1: 4 densities with equality constraints

% Algorithmic Parameters
nt=[1,1,1,1]*10^4;
s=length(nt); n=sum(nt);
lam=nt/n;

% Gibbs Sampling from f_s (Conditional on the rare-event of interest)
Xs = fs_rand(gamma,alpha,d,nt(4));

% Gibbs Sampling from f_3 
X3 = f3_rand(gamma,alpha,d,nt(3));

% Computing C_kj = max{0,gamma-sum_{j neq i}(X_ki)}
% Take a subsample of Xs to estimate product of marginals
m=nt(s)*0.5;
Xss=Xs(randperm(m),:);
C = NaN(m,d);
sums=sum(Xss,2);
for i=1:d
   C(:,i)=max(0,gamma-sums+Xss(:,i));
end

% Setup for efficient calculation of the product of marginals
C=sort(C(:),'ascend');
pi=cumsum(exp(C.^alpha))/(m*d);

% Sampling from f_2 (Reference density)
X2 = f2_rand(C,alpha,d,nt(2));

% Sampling from f_1 (Reference density)
X1 = f1_rand(gamma,alpha,d,nt(1));

% Pooled Samples
X = [X1; X2; X3; Xs];

% Coefficients in objective function D(z)
global Coeff
Coeff=NaN(4,n);
Coeff(1,:)=sum(X>=gamma,2); 
Coeff(3,:)=sum(X,2)>=gamma & max(X,[],2)<gamma;
Coeff(4,:)=sum(X,2)>=gamma; 

% Coefficients from f2 (efficient implementation)
[~,bin]=histc(X,C);
bin(bin==0)=size(C,1);
Coeff(2,:) = prod(pi(bin),2)';

% Optimization Options
options = optimset('Algorithm','interior-point','TolX',10^-14,'TolFun',10^-10,...
                   'GradConstr','on','Display','iter-detailed','GradObj','on',...
                   'Hessian','user-supplied','HessFcn',@hessianfcn,...
                   'MaxFunEvals',10^5,'MaxIter',10^5); 
               
% Constrained Optimization with fmincon              
[z,fval,exitflag,output] = fmincon(@(z)myfun(z,lam),zeros(1,4),...
    [],[],[],[],[],[],@(z)mycon(z,lam,-gamma^alpha+log(d)),options);

% Extract ell from optimal z
ell_con=lam.*exp(-z);

% Retrieve ELM estimator using analytical knowledge of ell1
prob=exp(-gamma^alpha)*d;
ell_con=ell_con/ell_con(1)*prob;

ell_con

%% MCIS Estimator
N=n;
ell=NaN(N,1);
X_IS = f2_rand(C,alpha,d,N);

% Efficient Implementation
[dummy,binIS]=histc(X_IS,C);
binIS(binIS==0)=size(C,1);
ell = (sum(X_IS,2)>gamma)./prod(pi(binIS),2);
ell_is=mean(ell); 
RE_is=std(ell)/mean(ell)/sqrt(nt(2));
[ell_is, RE_is]

%% Asmussen-Kroese conditional estimator
N=n;
Y=NaN(N,1);
for i=1:N
    x=(-log(rand(1,d-1))).^(1/alpha);
    Y(i)=d*exp(-max([x,gamma-sum(x)])^alpha);
end
ell_cond = mean(Y); 
RE_cond = std(Y)/mean(Y)/sqrt(N);
[ell_cond, RE_cond]
\end{Verbatim}

\subsection{Functions for Sampling}

\begin{Verbatim}[fontsize=\footnotesize,frame=topline,label=Function {\tt f1\_rand.m} to sample from $f_1(x)$]

%%%
% Sampling from f1: main 'heavy-tailed' component
%%%
function out=f1_rand(gamma,alpha,d,n)

out = NaN(n, d);
prob = ceil(rand(n,1)*d); 
for i = 1:n
   for j = 1:d       
       if ( j == prob(i))
        out(i,j) = (gamma^alpha -log(rand))^(1/alpha);
       else
        out(i,j) = (-log(rand))^(1/alpha);
       end       
   end
end

\end{Verbatim}

\begin{Verbatim}[fontsize=\footnotesize,frame=topline,label=Function {\tt f2\_rand.m} to sample from $f_2(x)$]

%%
% Sampling from f2: product of marginals from the zero-variance density
%%%
function out = f2_rand(C,alpha,d,n)

% generate j in (1,...,ns*d) with equal probability
J = ceil(rand([n,d]).*(size(C,1)));

out = ((C(J)).^alpha-log(rand([n d]))).^(1/alpha);   

\end{Verbatim}

\begin{Verbatim}[fontsize=\footnotesize,frame=topline,label=Function {\tt f3\_rand.m} to sample from $f_3(x)$]

%%%
% Gibbs Sampling from f3: residual component
%%%
function out=f3_rand(gamma,alpha,d,n)

x=ones(1,d)*gamma/d;
s=sum(x); out=NaN(n,d);
for i=1:n
    for j=1:d
        s=s-x(j);
        x(j)=trwbl(max(gamma-s,0),gamma,alpha);
        s=s+x(j);
    end
    
    out(i,:)=x(randperm(d));
    
end

\end{Verbatim}

\begin{Verbatim}[fontsize=\footnotesize,frame=topline,label=Function {\tt trwbl.m} to sample from truncated Weibull distribution]

%%%
% Simulating from general truncated Weibull(alpha,1) distribution
%%%
function out=trwbl(a,b,alpha)

out=a.^alpha-log(1-rand(size(a)).*(1-exp(a.^alpha-b^alpha)));
out=out.^(1/alpha);

\end{Verbatim}

\begin{Verbatim}[fontsize=\footnotesize,frame=topline,label=Function {\tt fs\_rand.m} to sample from $f_s(x)$]

%%%
% Gibbs Sampling from f4: zero-variance density
%%%
function out=fs_rand(gamma,alpha,d,n)

x=ones(1,d)*gamma/d;
s=sum(x); out=NaN(n,d);
for i=1:n
    for j=1:d
        s=s-x(j);
        x(j)=(-log(rand)+(max(0,gamma-s))^alpha)^(1/alpha);
        s=s+x(j);
    end    
    out(i,:)=x(randperm(d));    
end

\end{Verbatim}

\subsection{Functions for Optimization}

\begin{Verbatim}[fontsize=\footnotesize,frame=topline,label=Function {\tt myfun.m} to evaluate the objective function $D(z)$]

%%%
% Objective function
%%%
function [fval,gradval]=myfun(z,lam)

global Coeff
z=z(:)';
[s,n]=size(Coeff);
y=exp(z)*Coeff;
fval=sum(log(y))/n-lam*z';

if nargout>1
    gradval=exp(z').*sum(Coeff./repmat(y,s,1),2)/n-lam';
end

\end{Verbatim}

\begin{Verbatim}[fontsize=\footnotesize,frame=topline,label=Function {\tt mycon.m} to specify optimization constraints]

%%%
% Function to specify constraints
%%%
function [c,ceq,gradc,gradceq] = mycon(z,lam,p)

% 4 densities
% 2 constraints
c=[];
ceq=[sum(lam.*z); 
    log(lam(1))-z(1)+z(2)-log(lam(2))-p];

if nargout > 2
    gradc = [];
    gradceq = [lam(1),lam(2),lam(3),lam(4);
              -1,1,0,0]';
end

\end{Verbatim}

\begin{Verbatim}[fontsize=\footnotesize,frame=topline,label=Function {\tt hessianfcn.m} to compute Hessian of objective function ]

%%%
% Compute Hessian of objective function
%%%
function out=hessianfcn(z,Lagr_mult)
global Coeff
z=z(:)';
[s,n]=size(Coeff);
B=exp(z)*Coeff;
y=Coeff./repmat(B,s,1);

out=diag(exp(z').*sum(y,2))-(y*y').*(exp(z')*exp(z));

out=out/n;

\end{Verbatim}

\section{ELM scheme B: 2 densities with inequality constraint}
\label{appendix:matlab_ELM_scheme2}

The main program {\tt weibull\_main\_B.m} computes the estimator for ELM scheme B.
The lower bound probability is evaluated by calling {\tt lowerbound.m}. 
Functions used to sample from $f_2(x)$ and to perform the constrained optimization are also provided below.
This is illustrated in figure~\ref{fig:functions_ELM2} below.

\begin{figure}[H]
\centering
\caption{Functions for ELM Scheme B} \label{fig:functions_ELM2}
\begin{tikzpicture}[%
  grow via three points={one child at (0.5,-0.7) and
  two children at (0.5,-0.7) and (0.5,-1.4)},
  edge from parent path={(\tikzparentnode.south) |- (\tikzchildnode.west)}]
  \node { weibull\_main\_B}
    child { node {lowerbound}
    	  child { node {q}
                  child { node {convolution}} 
          }
    }
    child [missing] {}	
    child [missing] {}
    child { node {f2\_rand}}
    child { node {fmincon}
      child { node {myfun}}
      child { node {mycon}}
      child { node {hessianfcn}}
    };
\end{tikzpicture}
\end{figure}
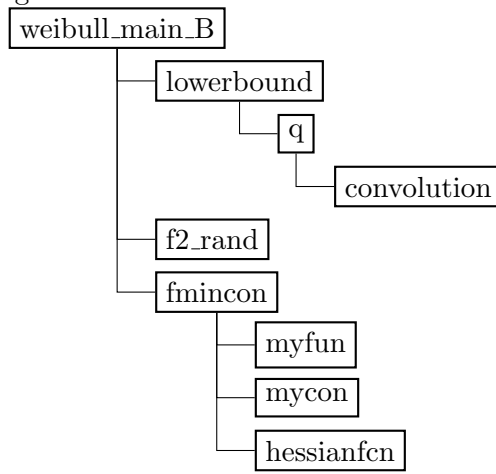

\subsection{Main program}

\begin{Verbatim}[fontsize=\footnotesize,frame=topline,label= Main program {\tt weibull\_main\_B.m }]

%%%
% weibull_main_B.m                                                
% Date: 23/10/2013                                                                        
% Description: Main program to compute the lower bound probability and
% estimator from ELM scheme B
%%%


format long g
format compact
clear all
clc

% Problem Parameters
gamma=90;
d=10;
alpha=0.9;

% Calculate optimal lambda, lower bound ell_1 proability
[prob,lamstar] = lowerbound(gamma,d,alpha);

% Calculate gamma star
gamstar = alpha*gamma + (1-alpha)*sum(lamstar.^(1/alpha));


% Algorithmic parameters
nt=[.1,1]*10^5;
s=length(nt); n=sum(nt);
lam=nt/n;

% Gibbs Sampling from f_2 zero variance density
% Conditional on the rare-event of interest
X2 = f2_rand(gamma,alpha,d,nt(2));

global Coeff
Coeff=NaN(2,n);
Coeff(1,:) = [ones(1,nt(1)) 
             (sum(repmat(lamstar.^(1/alpha-1),nt(2),1).*X2,2)>= gamstar)'];
Coeff(2,:) = ones(1,n);

% Optimization Options
options = optimset('Algorithm','interior-point','TolX',10^-14,'TolFun',10^-10,...
                   'GradConstr','on','Display','iter-detailed','GradObj','on',...
                   'Hessian','user-supplied','HessFcn',@hessianfcn,...
                   'MaxFunEvals',10^5,'MaxIter',10^5); 
% Constrained Optimization               
[z,fval,exitflag,output] = fmincon(@(z)myfun(z,lam),zeros(1,2),...
    [],[],[],[],[],[],@(z)mycon(z,lam),options);

ell_con=lam.*exp(-z);
ell_con=ell_con/ell_con(1)*prob;
ell_con

\end{Verbatim}

\subsection{Functions for Lower Bound Probability}

\begin{Verbatim}[fontsize=\footnotesize,frame=topline,label= Function {\tt lowerbound.m} to calculate $\ell_{L}$ and optimal $\boldsymbol\lambda^*$]

%%%
% Solves the nonlinear optimization using cross entropy optimization
% to find lower bound probability and optimal lambda 
%%%
function [prob lamstar] = lowerbound(gamma,d,alpha) 

se=3*ones(1,d); m=zeros(1,d); S_old=inf;
N=1000;
for iter=1:10^6
L=randn(N,d).*repmat(se,N,1)+repmat(m,N,1);
  for i=1:N
  S(i)=q(L(i,:)',gamma,alpha);
  end
  I=S>prctile(S,50);
  m=mean(L(I,:));
  se=std(L(I,:));
  if norm(max(S)-S_old)/max(S)<10^-6, break, end
  S_old=max(S);
end

lamstar = exp(m);
prob = q(m,gamma,alpha);

\end{Verbatim}

\begin{Verbatim}[fontsize=\footnotesize,frame=topline,label= Function {\tt q.m}]

%%%
% Function to compute the tail probability of Generalized Erlang distribution 
%%%
function out=q(lam,gamma,a)

d=length(lam);
beta = NaN(1,d);
for j=1:d
    beta(j)=sum(exp(lam(j:d)*(1/a-1)))/(d-j+1);
end

out=convolution(a*gamma+(1-a)*sum(exp(lam/a)),1./beta);
 
\end{Verbatim}

\begin{Verbatim}[fontsize=\footnotesize,frame=topline,label= Function {\tt convolution.m}]

%%
% Function to calculate the convolution P(A_1+...+A_b>t)
%%
function ell=convolution(t,nu)
b=length(nu);
A=diag(-nu)+diag(nu(1:b-1),1);
A=expm(A*t);
ell=sum(A(1,:));
ell=ell*(ell<1)*(ell>0);

\end{Verbatim}

\subsection{Functions for Sampling}

\begin{Verbatim}[fontsize=\footnotesize,frame=topline,label= Function {\tt f2\_rand.m} to sample from zero-variance density]

%%%
% Gibbs sampling from f2: zero-variance density
%%%
function X=f2_rand(gamma,alpha,d,n)

x=ones(1,d);
s=sum(x.^(1/alpha)); X=NaN(n,d);
for i=1:n
    for j=1:d
        s=s-x(j)^(1/alpha);
        x(j)= -log(rand) + max(0,(gamma-s))^alpha;   
        s=s+x(j)^(1/alpha);
    end
    
    X(i,:)=x(randperm(d));
    
end
X = sort(X,2);

\end{Verbatim}

\subsection{Functions for Optimization}

\begin{Verbatim}[fontsize=\footnotesize,frame=topline,label= Function {\tt myfun.m} to evaluate the objective function $D(z)$]

%%%
% Objective function
%%%
function [fval,gradval]=myfun(z,lam)
global Coeff
z=z(:)';
[s,n]=size(Coeff);
y=exp(z)*Coeff;
fval=sum(log(y))/n-lam*z';

if nargout>1
    gradval=exp(z').*sum(Coeff./repmat(y,s,1),2)/n-lam';
end

\end{Verbatim}

\begin{Verbatim}[fontsize=\footnotesize,frame=topline,label= Function {\tt mycon.m} to specify optimization constraints]

%%%
% Function to specify constraints
%%%
function [c,ceq,gradc,gradceq] = mycon(z,lam)

c=z(2)-z(1)-(log(lam(2))-log(lam(1)));
ceq=sum(lam.*z);
 
if nargout > 2
    gradc = [-1;1]; 
    gradceq = [lam(1);lam(2)];  
end

\end{Verbatim}

\begin{Verbatim}[fontsize=\footnotesize,frame=topline,label= Function {\tt hessianfcn.m} to compute Hessian of objective function]

%%%
% Compute Hessian of objective function
%%%
function out=hessianfcn(z,Lagr_mult)
global Coeff
z=z(:)';
[s,n]=size(Coeff);
B=exp(z)*Coeff;
y=Coeff./repmat(B,s,1);

out=diag(exp(z').*sum(y,2))-(y*y').*(exp(z')*exp(z));

out=out/n;

\end{Verbatim}

\end{appendices}

\clearpage																			
\addcontentsline{toc}{chapter}{References}      
\bibliographystyle{plain}
\bibliography{main}

\end{document}